\documentclass[a4paper,twocolumn,11pt,unpublished]{quantumarticle}
\pdfoutput=1
\usepackage[utf8]{inputenc}
\usepackage[english]{babel}
\usepackage[T1]{fontenc}
\usepackage{amsmath,amssymb,bm}
\usepackage{hyperref}
\usepackage{orcidlink}
\usepackage{graphicx}
\usepackage{xcolor}
\usepackage{booktabs}
\usepackage[numbers,sort&compress]{natbib}
\hypersetup{colorlinks=true,linkcolor=blue,citecolor=blue,urlcolor=blue}

\newcommand{\Ra}{R_a}
\newcommand{\Pop}{P_{\rm opt}}
\newcommand{\RGW}{R_a^{\rm GW}}
\newcommand{\us}{\,\mu\mathrm{s}}

\begin{document}

\title{Analog quantum approximate optimization algorithm for solving the weighted MaxCut problem on three-dimensional single-atom Rydberg array}

\author{Junxi Chen~\orcidlink{0009-0001-2107-2543}}
\affiliation{Rzhanov Institute of Semiconductor Physics SB RAS, 630090 Novosibirsk, Russia}
\affiliation{Novosibirsk State University, 630090, Novosibirsk, Russia}

\author{I. I. Beterov~\orcidlink{0000-0002-6596-6741}}
\affiliation{Rzhanov Institute of Semiconductor Physics SB RAS, 630090 Novosibirsk, Russia}
\affiliation{Novosibirsk State University, 630090, Novosibirsk, Russia}
\affiliation{Novosibirsk State Technical University, 630073, Novosibirsk, Russia}
\affiliation{Institute of Laser Physics SB RAS, 630090, Novosibirsk, Russia}

\begin{abstract}
We theoretically investigate an analog implementation of the quantum approximate optimization algorithm (QAOA) for the weighted  maximum cut problem (MaxCut) problem. The protocol uses a three-dimensional array of individually trapped Rydberg atoms driven by globally addressed laser pulses, with site-dependent detunings applied near the Rydberg resonance. 
As benchmarks, we consider twenty randomly weighted 3-regular graphs with 12 vertices each. Problem weights are encoded directly in the interatomic interaction energies by choosing appropriate three-dimensional atom positions.  We compare two parameter-optimization strategies: discretized quantum annealing and warm-start initialization using a graph neural network (GNN). The combined optimization strategy creates a portfolio that outperforms the Goemans–Williamson (GW) rounding almost in all instances in the noiseless case. We analyze the sampling statistics and the total pulse-time cost of both approaches. Finally, we emulate the performance of the optimized schedules on quantum trajectories using a realistic error model of the Aquila neutral-atom quantum processor.\end{abstract}

\maketitle

\section{Introduction}
\label{sec:intro}

The rapid development of noisy intermediate-scale quantum (NISQ) processors~\cite{Preskill2018} in recent years has greatly increased the interest in hybrid quantum-classical  computation approaches to a number of NP-hard problems. The weighted MaxCut problem is a fundamental combinatorial problem which has applications in very-large-scale integration  circuit design, data clustering, and network infrastructure layouts. The problem is APX-hard~\cite{Papadimitriou1991}, i.e., unless P${}={}$NP there is no polynomial-time algorithm that arbitrarily approximates the optimum. 

The family of quantum approximate optimization algorithms (QAOA) has recently emerged to solve combinatorial problems with potential runtime speedups~\cite{Farhi2014,Hadfield2019}. QAOA alternates the
evolution under a diagonal cost Hamiltonian with the evolution under a non-commuting mixer Hamiltonian, which is required to leave the local optima and to investigate the whole parameter space. The optimization scheme consists of $p$ subsequent pairs of cost and mixer optimization layers, where $p$ is the optimization depth. The application of QAOA to the MaxCut problem on regular graphs has already been proposed~\cite{Farhi2014,Hadfield2019,Wurtz2021PRA,Basso2022,Wybo2025,Li2025LowGirth}. QAOA has been shown to approach the adiabatic solution at large depths~\cite{Farhi2014,Zhou2020,Sack2021,Wurtz2022Counter},
while its limitations at fixed depths $p$ have also been identified. QAOA is a local algorithm, which imposes additional bounds for  high-girth graphs~\cite{Farhi2020Typical,Marwaha2021,Barak2022}. In bipartite-like instances, the solution is limited by the symmetry protection~\cite{Bravyi2020}. Classical optimization becomes hard in the worst case of plain landscapes, known as Barren plateaus~\cite{Bittel2021}. It also suffers from vanishing gradients in many settings~\cite{McClean2018,Larocca2025}. These findings were summarized in reviews~\cite{Blekos2024,Abbas2024}.

The analog version of QAOA relies on natural properties and dynamics of the quantum system, instead of the use of the compiled quantum gate sequences. Neutral atoms excited to Rydberg states are a promising platform for the analog versions of
QAOA~\cite{Saffman2010,Browaeys2020,Henriet2020,Morgado2021}. The van der Waals interaction
provides a programmable two-body Ising coupling. The global and local laser drivings  provide the mixer and single-site fields, and the evolution can be programmed as a sequence of laser pulses ~\cite{Pichler2018,Zhou2020,Ebadi2022,Wurtz2023Aquila}. Experiments  on Rydberg 
arrays have already addressed combinatorial and optimization problems, such as finding the maximum independent set on unit-disk
graphs~\cite{Ebadi2022,Kim2022Wires,Byun2022} and gate-based QAOA for
MaxCut problem~\cite{Graham2022}. The weighted variants of these problems can be solved through local light 
shifts and individual addressing in atomic arrays~\cite{Farouk2024,deOliveira2026,Goswami2024}. The individual addressing is also of interest for investigation of quantum spin liquids~\cite{Farouk2026}.

The weighted MaxCut problem in graph theory is defined as follows~\cite{Karp1972}. Let  $G=(\mathcal V,E,w)$ be a graph on a set of $L$ vertices  $\mathcal V=\{1,...,L\}$ with a set of edges $E\subset \mathcal V^2$ and a weight function $w:\,E\Rightarrow \mathbb{R}$ that assigns a non-negative weight $w_{ij}$ to each edge $(i, j) \in E$ (with $w_{ij} = 0$ if $(i, j) \notin E$). This problem consists in finding a partition $\mathcal V = \mathcal V_0 \cup \mathcal V_1$ with $\mathcal V_0 \cap \mathcal V_1 = \emptyset$ that maximizes the total weight of the edges $(i, j) \in E$ that connect the two subsets of vertices, namely $i \in \mathcal V_0$ and $j \in \mathcal V_1$:
$\max_{\mathcal V_0, \mathcal V_1} \sum_{i \in \mathcal V_0, j \in \mathcal V_1} w_{ij}$.

The efficient approximate solutions of NP-hard maximization problems are  characterized by the approximation ratio $R_a$, which is the ratio of the value obtained by the approximate algorithm to the exact optimum.
For solving the MaxCut problem, the natural classical reference is the
Goemans--Williamson (GW) algorithm~\cite{GW1995} with the worst-case guaranty approximation ratio of $R_a=0.878$. This algorithm is
optimal under the conjecture of unique games~\cite{Khot2007,Hastad2001}. However, in some instances, GW performs much better than its guaranty. For any alternative algorithm, a fair comparison must be made with what GW actually achieves on the same graphs, both for a single rounding and for repeated
sampling~\cite{Lykov2023,Dupont2022}. That is what we compare in this work.

Embedding arbitrary graphs into the geometric interaction pattern requires either the use of so-called gadgets~\cite{Nguyen2023,Lanthaler2023,
Lechner2015,Glaetzle2017,Dlaska2022}. Their overhead and interpretability have  been  critically analyzed recently~\cite{Cazals2025,deCorrec2025}. In contrast, we consider direct placement of atoms in such a way that the interatomic
coupling energies reproduce the target weights. Bauer \emph{et al.}~\cite{Bauer2026} showed that the weighted MaxCut problem can be solved with atomic interactions treated as weights, also using local detunings from the resonance at Rydberg excitation. The latter allows canceling the single-site terms in the optimization Hamiltonian, which arise due to the form how the van der Waals interaction appears in the Hamiltonian. In this case, the zero-drive evolution is reduced to a pure Ising evolution. This approach was initially considered  only for planar graphs derived from square and honeycomb
lattices~\cite{Bauer2026}. 

We generalize it for random weighted three-regular graphs, and place the atoms in three dimensions~\cite{Barredo2018,Kumar2018,Shi2020,Shah2025}. In this case the couplings of graph edges encode the edge weights, while the non-edge interactions are suppressed, but not absent. The same rule, strong couplings on the graph edges and weak couplings elsewhere, was recently used to encode molecular graphs in two-dimensional registers, where the measured correlations served as structural features for graph neural networks~\cite{Garrigues2026}. The weights of our test graphs are drawn at random (Sec.~\ref{sec:instances}), but the general approach is not tied to random weights. Any set of positive weights can be encoded, provided the geometry can realize it, i.e., the target distances $\propto w_{ij}^{-1/6}$ admit a three-dimensional placement in which every non-edge pair of atoms is farther apart than every edge pair.

This condition bounds the attainable spread of the weights; in our instances, the largest and smallest edge weights may differ by a factor of around $\sim 3$. Our emulation keeps every pairwise coupling in the array. The non-edge couplings, which any real atomic array produces, are considered as part of the system dynamics, rather than neglected.

The annealing structure of  the optimal QAOA schedules was initially established for
gate-model circuits, which have a pure transverse-field mixer~\cite{Zhou2020}. In this paper, we demonstrate that for the analog laser pulse sequence, where the mixer invariably incorporates the cost Hamiltonian $H_C$, expressed through atomic interactions, a layer-wise optimizer cannot intrinsically discover such an annealing-like schedule. However, imposing a single constraint on the duration of the cost segments successfully guides the optimizer toward the required annealing schedule. We validated that by tracking the ground state of each layer's generator.

We use warm starts from classical
relaxations~\cite{Egger2021,Tate2023Quantum,He2026RWS} and from GNN-based
initializations~\cite{Jain2022}, which  were initially formulated for gate implementations of QAOA. We translate the warm start parameters into the
site-resolved Rabi amplitudes and detunings, and identify the sign of the longitudinal mixer field,
for which the warm state is stationary in the hardware convention.

Beyond the approximation ratio, we compare our protocol with GW on the same instances both in expectation values and in sampling~\cite{Lykov2023,Dupont2022}, and quantify the
optimization cost,  with and without parameter transfer. We evaluate the optimized schedules by the
trajectory simulation, using the error model of a neutral-atom quantum processor. That shows which of the conclusions, made for the noiseless case, remain valid in the case of commonly considered error rates.

Three questions arise once such an analog protocol is optimized at depth. The first concern is the approximation ratio
landscape. Even the depth-one objective of a physical Rydberg sequence is multimodal, and the
choice of basin at small depth is not innocuous. Zhou \emph{et al.}~\cite{Zhou2020} showed that for the
gate-model QAOA the optimal schedules resemble discretized annealing. This structure can
be exploited through interpolation. Then, several works initialized QAOA from the annealing
paths~\cite{Sack2021,Brady2021,Wurtz2022Counter}. We examine if the optimizer finds this annealing-like structure in a specific case of the analog Rydberg sequence, in which the interatomic interactions are always present, in contrast to commonly considered gate-based optimization schemes.

The second question is the use of warm starts. It has been shown that starting QAOA from a biased product state, built from a classical relaxation, can beat the GW algorithm at low
depths~\cite{Egger2021,Tate2023Quantum,Tate2023TQC,He2026RWS}. Graph neural networks
(GNNs) can supply such marginals cheaply~\cite{Jain2022,Schuetz2022}, with the caveats that GNN
heuristics are not stronger than simple classical algorithms on their
own~\cite{Boettcher2023,Angelini2023}. However, a warm start can trap QAOA near the classical string from which it started ~\cite{Cain2022}. In an analog setting, the custom mixer of
Ref.~\cite{Egger2021} becomes a site-resolved drive, and its sign has to be determined in the conventions of the hardware Hamiltonian. 

The third question is the cost. Each objective evaluation
is an experiment, and the cost of training QAOA parameters from scratch for each instance is a
practical barrier~\cite{Tibaldi2023,Otterbach2017,Verdon2019,Khairy2020}. The optimal parameters concentrate and transfer across the instances~\cite{Brandao2018,Galda2021,Wurtz2021Fixed,
Shaydulin2023Transfer,Sureshbabu2024,Montanez2025,Xu2025GAT,Jiang2025QSeer}, which suggests that most of that cost can be avoided.

This paper answers these questions for a sample of twenty weighted three-regular graphs on 12 vertices,
embedded in 3D and emulated directly with all pairwise couplings taken into account. We find the following: (i) a layer-by-layer
optimizer gets trapped on a non-adiabatic branch. A single constraint on the cost-segment duration moves it onto an annealing branch with expected cut exceeding that of GW rounding at
$p\gtrsim16$. (ii) The GNN warm start requires the mixer sign for which the warm state is
stationary. (iii) The cold start and the GNN warm start fail on different instances, which makes a two-member portfolio strictly better than either.

The paper is organized as follows.
Sec.~\ref{sec:model} defines the physical model, the graph embedding in 3D Rydberg arrays,
the GNN warm start and the figures of merit. Sec.~\ref{sec:omega}  discusses the role of the mixer
Rabi frequency. Sec.~\ref{sec:branches} analyzes the landscape of the approximation ratio and the optimized pulse schedules.
Sec.~\ref{sec:warm} describes the  GNN warm start in detail. Sec.~\ref{sec:gw} compares with GW and with the portfolio,
Sec.~\ref{sec:sampling} provides the sampling statistics, and Sec.~\ref{sec:cost} considers
the optimization cost. Sec.~\ref{sec:noise} states the control requirements and evaluates
the algorithm using the error model of a neutral-atom quantum processor. Finally, Sec.~\ref{sec:discussion} discusses the
scope of the results. 

\section{Model, instances and methods}
\label{sec:model}

\subsection{Rydberg Hamiltonian and laser pulse sequence}
\label{sec:hamiltonian}

\begin{figure*}[!t]
\centering
\includegraphics[width=\textwidth]{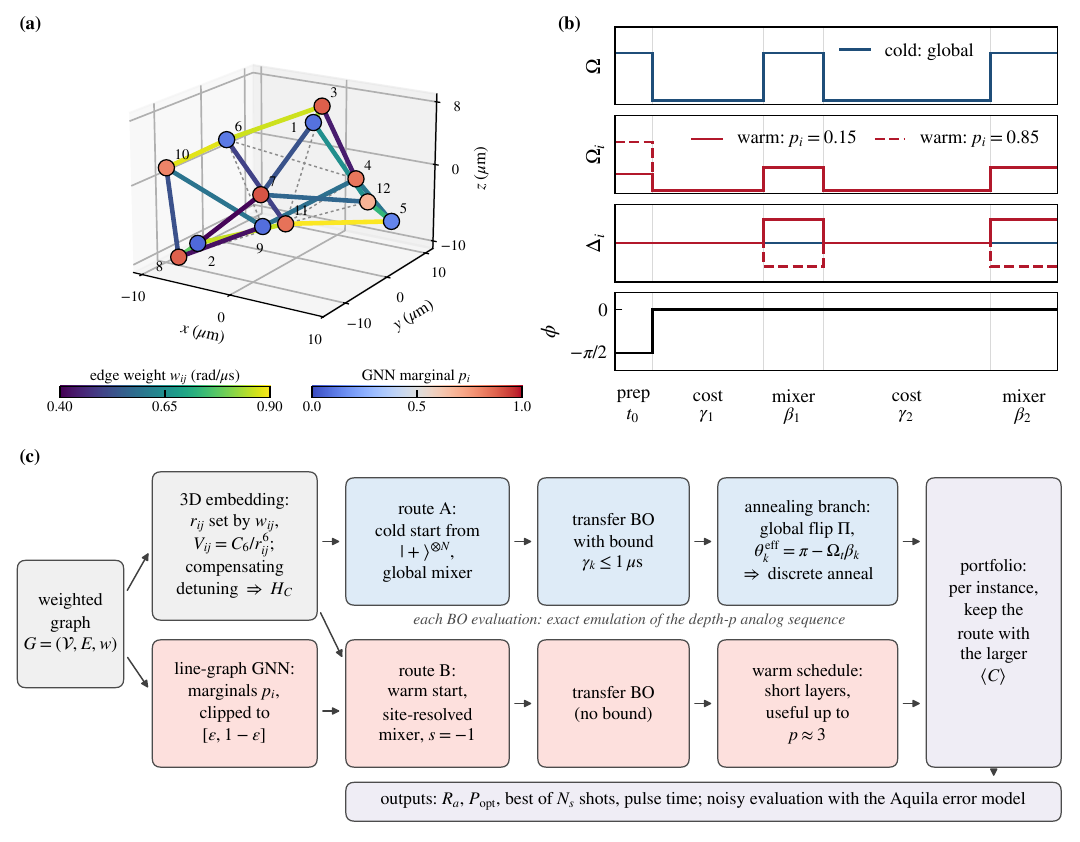}
\caption{Problem setting. (a) Graph instance 1 embedded in 3D atomic array. The numbers label the atoms, which are the vertices of graph $G$. Solid lines represent 18 edges of $G$, colored by the edge weight
$w_{ij}=V_{ij}/4$ (left-hand color bar; $0.40$--$0.90$~rad/$\mu$s in this instance, values listed in
Table~\ref{tab:inst1}). Dashed gray lines show 8 strongest couplings between the atoms representing non-adjacent graph vertices with $V_{ij}/4=0.12$--$0.26$~rad/$\mu$s. These couplings are not part of $G$ MaxCut problem, but still exist in the Hamiltonian from Eq.~\eqref{eq:HC}. Node color indicates GNN marginals $p_i$ (right-hand color bar), which are the probabilities assigned by
the network to vertex $i$ lying on the side of the cut mapped to $|r\rangle$. The rounding
($p_i>1/2$) places  red-colored atoms in $|r\rangle$ and blue-colored atoms in $|g\rangle$. (b)~Pulse sequence for $p=2$ (durations not to scale): global Rabi frequency $\Omega$ (cold
start, blue), site-resolved $\Omega_i$ (warm start, red), detuning $\Delta_i$ and laser phase
$\phi$. For warm start, two atoms are shown as an example: the first with $p_i=0.15$ (solid) and the second with
$p_i=0.85$ (dashed), with mixer amplitudes from Eqs.~\eqref{eq:prep}--\eqref{eq:mix}.
$\Delta_i$ is the compensation detuning $\frac12\sum_jV_{ij}$ from Eq.~\eqref{eq:comp}, shifted by
$s\,\Omega_tq_i$ in the warm-start mixer. (c) Workflow. The weighted graph is encoded once: the atom positions implement the weights
$w_{ij}=V_{ij}/4$. With the compensation detuning terms $\Delta_i$  of Eq.~\eqref{eq:comp}, the zero-drive
Hamiltonian is the Ising cost Hamiltonian $H_C$ used by both optimization routes. Route A (cold start, blue boxes)
begins from $|+\rangle^{\otimes N}$. The transfer optimizer with the bound $\gamma_k\le1\,\mu$s on
every cost segment yields the annealing branch:
after the global spin flip $\Pi$, its near-$\pi$ mixer rotations become small rotations
$\theta^{\rm eff}_k=\pi-\Omega_t\beta_k$ of a discretized anneal from the mixer to $H_C$
(Sec.~\ref{sec:branches}). In route B (warm start, red boxes) the GNN marginals $p_i$ set the initial
product state and the site-resolved mixer with $s=-1$ (Secs.~\ref{sec:gnn} and \ref{sec:warm});
its useful layers end at $p\approx3$. Every evaluation in the optimizer is the exact emulation of
the depth-$p$ analog sequence. The portfolio keeps, for each instance, the route with the larger
$\langle C\rangle$ (Sec.~\ref{sec:gw}). }
\label{fig:schematic}
\end{figure*}

In our model we consider a 3D array of $N$ individually addressed ultracold atoms, as shown in Fig.~\ref{fig:schematic}(a). The ground $|g\rangle$ and Rydberg  $|r\rangle$  states of the $i^{th}$ atom are coupled by a laser field with site-dependent Rabi frequency $\Omega_i(t)$ and detuning $\Delta_i(t)$, and a common time-dependent phase $\phi(t)$. In the rotating frame ($\hbar=1$, time in $\mu$s,
couplings in rad/$\mu$s) the system Hamiltonian is 
\begin{align}
H(t)={}&\sum_i\frac{\Omega_i(t)}{2}\bigl(e^{i\phi(t)}|g\rangle\langle r|_i+{\rm h.c.}\bigr)\nonumber\\
&-\sum_i\Delta_i(t)\,n_i+\sum_{i<j}V_{ij}\,n_in_j ,
\label{eq:H}
\end{align}
where $i,j=1,\dots,N$ label the atoms, ``h.c.'' denotes the Hermitian conjugate, $n_i=|r\rangle
\langle r|_i$ is the Rydberg atom number operator and
$V_{ij}=C_6/r_{ij}^6$ is the van der Waals interaction energy of atoms $i$ and $j$ at the interatomic distance
$r_{ij}=|\bm r_i-\bm r_j|$, with atom positions $\bm r_i$. 

We consider $^{87}$Rb atoms with
$|g\rangle=|5S_{1/2}\rangle$ and $|r\rangle=|70S_{1/2}\rangle$, coupled by a two-photon transition
through the off-resonant intermediate state $|6P_{3/2}\rangle$, as in the Aquila
device~\cite{Wurtz2023Aquila}. The van der Waals coefficient of the Rb pair state
$|70S_{1/2},70S_{1/2}\rangle$ is $C_6/h=862.69$~GHz\,$\mu$m$^6$~\cite{Wurtz2023Aquila}.  This value of $C_6$ is the default of the
Bloqade.jl Rydberg simulator~\cite{Bloqade2023}.
For two
atoms in the same $nS_{1/2}$ state, the interaction is nearly isotropic~\cite{Saffman2010}, and the interaction energy
$V_{ij}$ depends only on the interatomic distance, but not on the orientation of $\bm r_i-\bm r_j$. That  makes a three-dimensional placement possible.

We use the Pauli operators
\begin{align}
X_i&=|g\rangle\langle r|_i+|r\rangle\langle g|_i,\nonumber\\
Y_i&=-i|g\rangle\langle r|_i+i|r\rangle\langle g|_i,\nonumber\\
Z_i&=|g\rangle\langle g|_i-|r\rangle\langle r|_i ,
\label{eq:pauli}
\end{align}
so that $n_i=(1-Z_i)/2$ and the driving term of Eq.~\eqref{eq:H} equals
$\tfrac{1}{2}\Omega_i(t)\left[X_i\cos\phi(t)-Y_i\sin\phi(t)\right]$. If the interaction part of the  Hamiltonian from Eq.~\eqref{eq:H} is expressed through $Z_i$ operators, it will contain  single-site terms $-\tfrac14\sum_{j\ne i}V_{ij}Z_i$. Following Ref.~\cite{Bauer2026}, we eliminate them by adding a  static single-site detuning for the $i^{th}$ atom, defined as
\begin{equation}
\Delta_i=\tfrac12\sum_{j\neq i}V_{ij},
\label{eq:comp}
\end{equation}
where the sum runs over \emph{all} other atoms. With $\Omega_i=0$ the Hamiltonian~\eqref{eq:H} is then, up
to a constant,
\begin{align}
H_C&=\sum_{(i,j)\in E}w_{ij}Z_iZ_j+\sum_{(i,j)\notin E}\frac{V_{ij}}{4}Z_iZ_j,\nonumber\\
w_{ij}&=\frac{V_{ij}}{4},
\label{eq:HC}
\end{align}

which is invariant under the global spin flip $\Pi=\prod_iX_i$. 

The Hamiltonian~\eqref{eq:HC} represents the target graph $G=(\mathcal V,E,w)$, with one vertex per atom ($|\mathcal V|=N$), edge set $E$ and edge weights
$w_{ij}$. The first term of Eq.~\eqref{eq:HC} is the weighted MaxCut Hamiltonian of graph $G$, and the
second term collects the couplings between the atoms that are not adjacent in $G$.

The atoms are initially prepared in state $|g\rangle^{\otimes N}$. Our optimization sequence begins with the global laser pulse with Rabi frequency
$\Omega_0=5\pi$~rad/$\mu$s, phase $\phi=-\pi/2$ and duration $t_0=\pi/(2\Omega_0)=0.1\us$, similar to Ref.~\cite{Bauer2026}. With $\phi=-\pi/2$, the
driving term equals $\frac{1}{2}\Omega_0\sum_iY_i$. The atomic interactions are present during this pulse
as well: the interaction terms of Eq.~\eqref{eq:H} together with the compensation detuning give the following part of the Hamiltonian:
\begin{equation}
\sum_{i<j}V_{ij}\,n_in_j-\sum_i\Delta_i n_i=H_C+\mathrm{const},
\label{eq:intterms}
\end{equation}
which contains both the edge and the non-edge couplings from Eq.~\eqref{eq:HC}. The full Hamiltonian of the system evolution under laser pulse is
$\frac{1}{2}\Omega_0\sum_iY_i+H_C$. 

To avoid Rydberg blockade at the preparation stage, the laser pulse should be short and intensive. The drive rotates every atom by
$\Omega_0t_0=\pi/2$, while $H_C$ acts on atom $i$ as the longitudinal field of magnitude 
$\sum_{j\neq i}V_{ij}/4$ at most, and imprints the phases of order
$t_0\sum_{j\neq i}V_{ij}/4=(\pi/2)\sum_{j\neq i}V_{ij}/(4\Omega_0)$. These phases are small, when the driving Rabi
frequency exceeds the interatomic couplings $\Omega_0\gg\sum_{j\neq i}V_{ij}$. 

In this limit,
$H_C$ can be neglected during the global laser excitation pulse. This resembles a hard-pulse approximation of digital-analog schemes, in which single-qubit pulses are applied on top of the always-on interaction~\cite{ParraRodriguez2020,Headley2022}. Ref.~\cite{Bauer2026} uses it with
$w_{ij}\ll\Omega_0$. The laser pulse drives the single-qubit rotation $e^{-i\pi\sum_iY_i/4}$, which
prepares the state $|+\rangle^{\otimes N}$ of the whole atomic array, with the single-atom state 
$|+\rangle=(|g\rangle+|r\rangle)/\sqrt2$.

For the instances which we consider, the coupling field is 
$\sum_{j\neq i}V_{ij}/4\approx2.3$~rad/$\mu$s, on average, a factor of about seven below $\Omega_0$,
which is  close to the largest Rabi frequency used in current experiments
($15.8$~rad/$\mu$s in Ref.~\cite{Wurtz2023Aquila}). The phases imprinted by $H_C$  are of order $0.2$~rad, which
is  not negligible. Therefore, the prepared state is not exactly equal to $|+\rangle^{\otimes N}$, while being close to it. We do not ignore this in the calculations: all results in the rest of the paper are computed with  full Hamiltonian. 

The laser pulse sequence and  the optimization workflow are illustrated in Fig.~\ref{fig:schematic}(b) and Fig.~\ref{fig:schematic}(c), respectively. After 3D atom placement we follow two optimization routes: cold start A and warm start B. Both routes use similar structure of laser pulse sequences, as shown in Fig.~\ref{fig:schematic}(b), but for warm start the pulse parameters are individually defined for each atom in the array.

The initial preparation pulse is followed by
$p$ optimization layers, each consisting of a cost segment of duration $\gamma_k$ ($\Omega_i=0$) and a mixer
segment of duration $\beta_k$ ($\Omega_i=\Omega_t$, $\phi=0$) [Fig.~\ref{fig:schematic}(b)]. The detuning of
Eq.~\eqref{eq:comp} remains on throughout, and the mixer Hamiltonian is written as
\begin{equation}
H_M=\frac{\Omega_t}{2}\sum_i X_i+H_C ,
\label{eq:HM}
\end{equation}
where $\Omega_t$ is the Rabi frequency of the mixer. $H_M$ is
a transverse-field Ising Hamiltonian rather than a pure transverse field. The set of $2p$ variable scheme parameters are the pulse durations, restricted to $\gamma_k\in[0,3\pi]\us$ and $\beta_k\in[0,2\pi/\Omega_t]$, in such a way
that the mixer rotation angle $\Omega_t\beta_k$ covers the entire interval $[0,2\pi]$. The total pulse duration is
$T=t_0+\sum_k(\gamma_k+\beta_k)$. A layer with $\gamma_k=\beta_k=0$ is the identity, and a
depth-$p$ protocol contains every shallower one. Unless stated otherwise, we use $\Omega_t=15.7$
rad/$\mu$s ($2\pi\times2.5$~MHz), which is the largest value considered in Sec.~\ref{sec:omega}.

\subsection{Graph instances and three-dimensional embedding}
\label{sec:instances}

For our simulation we considered  twenty test instances, which are random three-regular graphs on $n=12$ vertices ($|E|=18$). Random target weights $\tilde w_{ij}$ were drawn uniformly in the range $[0.5,1]$~rad/$\mu$s and then converted into target interatomic distances $\tilde r_{ij}=(C_6/4\tilde w_{ij})^{1/6}$, which are in the range ($10.5$--$11.8\,\mu$m). The exact atomic arrangement requires strong couplings $V_{ij}$ between the atoms, joined by an edge of the graph, and weak couplings between non-adjacent atoms. 

This is not straightforward to achieve for an arbitrary graph configuration. The exact atomic positions were then found by minimizing the auxiliary energy
\begin{equation}
U=\sum_{i<j}\frac{C_6}{4r_{ij}^6}+\sum_{(i,j)\in E}k_{ij}\,r_{ij}^2,\qquad
k_{ij}=\frac{3C_6}{4\tilde r_{ij}^{\,8}},
\label{eq:embed}
\end{equation}
over the $3N$ coordinates, starting from random positions (L-BFGS followed by Nelder--Mead). The
first term repels every pair of atoms, whether or not they are joined by a graph edge, and pushes
non-adjacent atoms apart. 

The second term pulls  two atoms belonging to every graph edge together like a
spring of zero rest length; its stiffness $k_{ij}$ is chosen such that the energy of an isolated
edge, $C_6/4r^6+k_{ij}r^2$, is minimal exactly at $r=\tilde r_{ij}$. The function $U$ only serves to
find the geometry of the atomic array and is not used for simulation of the system dynamics. A force-based placement with a similar aim was used in Ref.~\cite{Garrigues2026} for molecular graphs. There the register is two-dimensional and all edges have the same coupling, whereas here the edge lengths carry the weights and random three-regular graphs require the third dimension. The repulsion of the other atoms shifts the
edges away from their target lengths. Therefore, the weights in our graph instances slightly differ from the target ones. 

We
quantify how faithfully the obtained placement reproduces the intended weights by the Pearson correlation
coefficient between the target $\tilde w_{ij}$ and the realized $w_{ij}$ over the 18 edges of each instance. This correlation
is insensitive to a common rescaling of all weights, which only rescales the cost durations, and
measures whether the relative sizes of the weights survive in the embedding. It ranges from $0.76$ to
$0.99$ (mean value $0.94$), so the target weights are reproduced up to moderate distortions. This
coefficient is only a diagnostic tool: since all calculations below use the realized weights, a lower coefficient
value indicates the change in the weights of the solved graph, but not the validity of the solution.  We accepted the placement only  if every non-edge atomic pair is farther apart than every edge pair. 

The weights used
everywhere below are the couplings $w_{ij}=V_{ij}/4$ for the final geometry of the atomic array, which range from
$0.28$ to $1.01$~rad/$\mu$s (mean $0.68$). Edge lengths are $10.5$--$13.0\,\mu$m, the arrays
extend over about $40\,\mu$m, and the vertical extent is in the range  $19$--$34\,\mu$m. Therefore, the embedding
requires a three-dimensional trap array~\cite{Barredo2018,Kumar2018,Shi2020,Shah2025}. An example of this arrangement, denoted as instance~1, is shown in Figure~\ref{fig:schematic}(a).

The number  $n=N=12$ of graph vertices is equal to the number of atoms, and $|E|$ is the number of edges. Two quantities
describe how well the geometry of the atomic array represents the graph. The unwanted interaction between the atoms, which are not connected by the graph edge, is characterized by the ratio
\begin{equation}
\eta=\frac{\sum_{(i,j)\notin E}V_{ij}}{\sum_{(i,j)\in E}V_{ij}}
\label{eq:eta}
\end{equation}
(here both sums run over unordered pairs of atoms), which ranges from $0.056$ to $0.276$ (mean $0.135$).
Because $\eta$ is the average, we also compare, in each instance, the strongest coupling between
non-adjacent atoms with the weakest edge coupling, defined as
$\max_{(i,j)\notin E}V_{ij}/\min_{(i,j)\in E}V_{ij}$. 

The acceptance criterion bounds this
ratio by one, since it equals $(\max_E r_{ij}/\min_{\notin E}r_{ij})^6$. Across twenty studied instances it
ranges from $0.11$ to $0.90$ (median $0.24$). It exceeds $0.5$ in seven instances and reaches $0.90$
in instance~12, where a single non-edge coupling is almost as strong as the weakest edge. Second, we check whether the ground state of the full $H_C$ is an
optimal cut of $G$. It was found to be true for 17 out of 20 instances. For instances 6, 12 and 16 the ground state of
Eq.~\eqref{eq:HC} has a cut ratio of $0.973$, $0.978$ and $0.990$. In these particular three instances even
a perfect adiabatic evolution does not return the optimal cut.

\subsection{Figures of merit and classical reference}
\label{sec:fom}

Measuring the final state $|\psi\rangle$ in the $\{|g\rangle,|r\rangle\}$ basis gives a bit string
$z\in\{0,1\}^N$ ($z_i=1$ for $|r\rangle$) with the probability $\Pr(z)=|\langle z|\psi\rangle|^2$. The
cut value of $z$ is the total weight of the edges with end points lying on different sides,
\begin{equation}
C(z)=\sum_{(i,j)\in E}w_{ij}\,(z_i-z_j)^2 ,
\label{eq:cut}
\end{equation}
since $(z_i-z_j)^2=1$ if $z_i\neq z_j$ and $0$ otherwise.  Equivalently, $C$ is the eigenvalue of
$\sum_{(i,j)\in E}w_{ij}(1-Z_iZ_j)/2$ on $|z\rangle$. The maximum cut
$C_{\max}=\max_zC(z)$ is obtained by exhaustive enumeration of the $2^N$ bit strings, and
$\langle C\rangle=\sum_z\Pr(z)\,C(z)$. 

We calculate the approximation ratio $\Ra=\langle C\rangle/C_{\max}$
and the probability of measuring an optimal cut
\begin{equation}
\Pop=\sum_{z\in\mathcal Z_{\rm opt}}\Pr(z),
\label{eq:popt}
\end{equation}
where $\mathcal Z_{\rm opt}=\{z:\,C(z)=C_{\max}\}$ contains every optimal string together with its complement
$\bar z$ ($\bar z_i=1-z_i$), which has the same cut value. We also report the best expected cut among $N_s$ independent samples
$z^{(1)},\dots,z^{(N_s)}$ drawn from $\Pr(z)$,
$\langle\max_{N_s}C\rangle\equiv\mathbb E\bigl[\max_{1\le m\le N_s}C(z^{(m)})\bigr]$,
evaluated with Eq.~\eqref{eq:bestN}. All three
quantities are computed exactly from the calculated probability of the bit string $\Pr(z)$; they are the outputs of both routes in Fig.~\ref{fig:schematic}(c) (box ``measure''), and the portfolio of Fig.~\ref{fig:schematic}(c) compares the routes by $\langle C\rangle$ (Sec.~\ref{sec:gw}).

For comparison with GW algorithm,  we solve the MaxCut semidefinite program (SDP) for each instance with CVXPY library~\cite{cvxpy2016}
and draw $2\times10^4$ random hyperplanes. The mean of the resulting cut ratio, $\RGW$ is the
expected cut of a single rounding. Across the test set (range
$0.930$--$0.970$) it averages $0.9477$, against $1.020$ for the SDP bound. The same samples give $\Pop^{\rm GW}$
(mean $0.354$) and $\langle\max_{N_s}C\rangle^{\rm GW}$, the same quantity with the samples
$z^{(m)}$ given by independent GW roundings. 

The claim that our protocol exceeds GW means $\Ra>\RGW$, i.e., we compare the expectation values for one sample for each algorithm. Paired comparisons over the 20 instances use the two-sided Wilcoxon signed-rank test unless stated
otherwise, and $P$ always denotes the resulting $P$ value (not to be confused with the depth $p$). We also give win/loss counts and sign-test $P$ values, where they differ
qualitatively. Error bars are the standard errors over tested instances.

\subsection{Emulation and optimization}
\label{sec:methods}

All objective values are computed by Krylov propagation of Eq.~\eqref{eq:H} on the full
$2^{12}$-dimensional space using a Rydberg simulation package Bloqade.jl~\cite{Bloqade2023}. As an independent check, we have written an emulator that diagonalizes $H_M$ once per instance and applies the cost segments as
diagonal phases. It reproduces every optimized value reported below (600 instance--depth pairs)
to within $5\times10^{-7}$ (Appendix~\ref{app:emulator}); we use it for investigation of the landscapes, sampling
statistics and pulse-time analysis.

Two optimizers are used. The transfer BO described below is the optimizer of both routes in
Fig.~\ref{fig:schematic}(c), with the bound $\gamma_k\le1\us$ in route A; the layer-wise BO
serves only as a baseline and is not part of the workflow.

The \emph{layer-wise Bayesian optimization} (BO) baseline grows the circuit
one layer at a time: each new layer is found by a 300-evaluation Gaussian-process search over
$(\gamma_p,\beta_p)$ with the earlier layers fixed. The earlier layers are then revisited in two
block-coordinate sweeps, and the full vector is optmized with Nelder--Mead~\cite{NelderMead1965}.
The surrogate is a Gaussian process with a spectral-mixture kernel~\cite{Wilson2013}. A variant seeds 
polish with the linear-ramp schedule of Trotterized quantum annealing (TQA)~\cite{Sack2021}. This baseline reaches $p=5$ at about
$3\times10^4$ evaluations per instance.

The \emph{transfer BO} reuses optimized parameters from other instances, in
the spirit of Refs.~\cite{Shaydulin2023Transfer,Galda2021,Montanez2025} and of meta-learned
initial designs~\cite{Feurer2015,Wistuba2015}. For each target instance a leave-one-out source
database is formed from the other 19 instances. At every depth the new layer is optimized by a
Gaussian process with an automatic-relevance-determination (ARD) Matérn-5/2 kernel~\cite{Rasmussen2006,Snoek2012}
and only 30 evaluations, with initial design containing the coordinate-wise median of the
sources' optimal layers (with $\gamma$ rescaled by the ratio of mean edge couplings, as in
Ref.~\cite{Shaydulin2023Transfer}) and four high-density source optima selected by kernel
density. 

For $p\ge2$ the interpolated vector of Zhou \emph{et al.}~\cite{Zhou2020} (INTERP) is
offered as an alternative starting point at the cost of one evaluation. Each layer ends with two
block-coordinate sweeps (30 evaluations per earlier layer and sweep) and a 150-evaluation
Nelder--Mead optimization of the full vector. Appendix~\ref{app:transfer} gives the details.

The \emph{annealing-branch} variant
differs in one constraint: every cost segment is limited to $\gamma_k\le1\us$, and its source
database is built under the same constraint. The source databases are built once per instance
and condition (80 Gaussian-process and 40 Nelder--Mead evaluations per layer) and are not
counted in the evaluation numbers, which we report for a target instance.

\subsection{Graph-neural-network warm start}
\label{sec:gnn}

The warm start uses vertex marginals $p_i\in(0,1)$ which represent the probability, assigned by the network to
vertex $i$ lying on the side of the cut, which is mapped to $|r\rangle$ atomic state. The values $p_i$ are produced in a single forward pass by a
line-graph neural network~\cite{Chen2019LGNN}, trained without labels on a separate pool of
embedded instances with the relaxed MaxCut loss of Ref.~\cite{Schuetz2022}, as in
Ref.~\cite{Jain2022}. The network uses powers of the adjacency matrix up to $J=3$. It has $32$ features per
node and edge and $8$ layers; its inputs are the weighted vertex degrees $\sum_jw_{ij}$ and the edge
weights $w_{ij}$, and it was trained with Adam (learning rate $10^{-3}$, batches of 24 graphs) on
20 instances of every even size $n=12$--$24$, generated with the embedding procedure of
Sec.~\ref{sec:instances} from seeds disjoint from those of the test instances. Here the network is
fixed and only its output matters. 

Rounding the
marginals ($p_i>1/2$) gives cuts with mean ratio $R_a=0.955$. The rounded cut was found to be optimal for 10 of 
20 instances. The rounded output is therefore itself a classical heuristic, and we compare with it
in Sec.~\ref{sec:gw}. Following Egger \emph{et al.}~\cite{Egger2021}, we clip the marginals to
$[\varepsilon,1-\varepsilon]$, where $\varepsilon\in[0,1/2)$ is the clipping parameter, and write
$q_i=2p_i-1\in(-1,1)$. The preparation pulse in the warm start case becomes site resolved, with Rabi frequency
$\Omega_{0,i}$ of the laser field acting on atom $i$:
\begin{equation}
\Omega_{0,i}=\frac{2\arcsin\sqrt{p_i}}{t_0}.
\label{eq:prep}
\end{equation}
When the interactions ($H_C$) are neglected during the $0.1\us$ pulse, this pulse prepares the superposition state
$|\psi_i\rangle=\sqrt{1-p_i}\,|g\rangle+\sqrt{p_i}\,|r\rangle$. The mixer field uses Rabi frequency and modified detuning
\begin{align}
\Omega_{i}&=2\Omega_t\sqrt{p_i(1-p_i)},\nonumber\\
\Delta_i&\to\Delta_i+s\,\Omega_t\,q_i,\qquad s=\pm1 ,
\label{eq:mix}
\end{align}
where $\Omega_i$ is the mixer Rabi frequency for atom $i$, and $\Delta_i$ on the right-hand side of Eq.~\eqref{eq:mix} is the initial
compensation detuning taken from Eq.~\eqref{eq:comp}. The value $s$ defines the sign of the additional
longitudinal field of the optimization protocol.

Dropping $H_C$, i.e., the interaction terms together with the compensation detuning, the mixer Hamiltonian of atom $i$ is written, up to a
constant, as
$H_{M,i}=\tfrac12\bm h_i\cdot\bm\sigma_i$ with $\bm\sigma_i=(X_i,Y_i,Z_i)$ and
$\bm h_i=\Omega_t\bigl(2\sqrt{p_i(1-p_i)},\,0,\,s\,q_i\bigr)$, while the Bloch vector $\bm b_i=\langle\psi_i|\bm\sigma_i|\psi_i\rangle$ of state
$|\psi_i\rangle$ has three components
$\bm b_i=(2\sqrt{p_i(1-p_i)},0,-q_i)$.
The vectors $\bm b_i$ and $\bm h_i$ are parallel, in which case $|\psi_i\rangle$ is also a mixer eigenstate,
only for the sign $s=-1$, as in Ref.~\cite{Egger2021}. For $s=+1$ the angle $\alpha_i$ between $\bm b_i$ and $\bm h_i$ obeys the relation
$\cos\alpha_i=8p_i(1-p_i)-1$, and the
mixer rotates the warm state away. 

The smallest survival probability of the single-atom superposition state reached during the mixer stage is
\begin{equation}
S_i\equiv\min_t\bigl|\langle\psi_i|e^{-iH_{M,i}t}|\psi_i\rangle\bigr|^2=\bigl[1-8p_i(1-p_i)\bigr]^2 ,
\label{eq:survival}
\end{equation}
which vanishes at $p_i=(2\pm\sqrt2)/4$. Section~\ref{sec:warm} compares the cases of both mixer signs. Implementing Eqs.~\eqref{eq:prep}
and \eqref{eq:mix} requires site-resolved Rabi amplitudes in addition to site-resolved
detunings (Sec.~\ref{sec:noise}).

\section{The mixer Rabi frequency}
\label{sec:omega}

In the atomic array the interactions remain present during the mixer stage: the mixer Hamiltonian, Eq.~\eqref{eq:HM}, contains the full $H_C$. Therefore, both the edge and the non-edge couplings of Eq.~\eqref{eq:HC} act on the system while the transverse field rotates it.
 At fixed rotation angle
$\theta=\Omega_t\beta$, the duration of the mixer stage and, hence, the undesired phase accumulated on non-edge
pairs, scale as $V_{ij}\theta/\Omega_t$. Larger values of $\Omega_t$  help, especially for
instances with larger $\eta$. 

\begin{figure}[!t]
\centering
\includegraphics[width=\columnwidth]{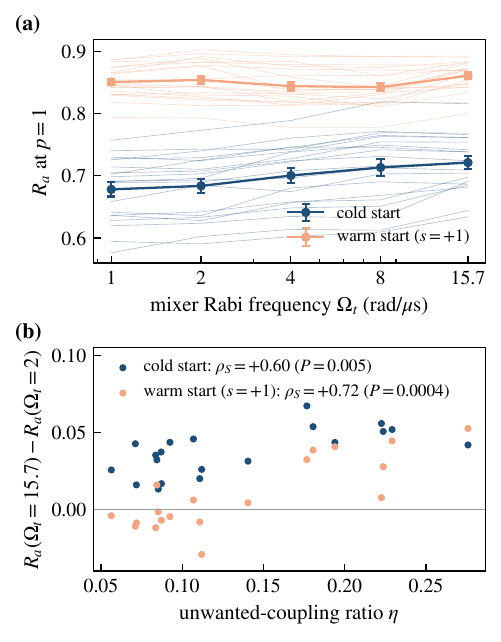}
\caption{ Dependence of the approximation ratio on the mixer Rabi frequency at $p=1$ (exact optimum of each instance).
(a) Mean $\Ra$ as a function of $\Omega_t$; thin lines represent  individual instances. (b) The difference 
$R_a(\Omega_t=15.7\, \mathrm{rad}/\mu\mathrm{s})-R_a(\Omega_t=2\, \mathrm{rad}/\mu\mathrm{s})$  as a function of  the unwanted-coupling ratio $\eta$ from Eq.~\eqref{eq:eta}.}
\label{fig:omega}
\end{figure}

Figure~\ref{fig:omega} confirms this statement at $p=1$, where the exact
optimum was obtained on a $60\times60$ grid, followed by local refinement. For cold start, the
mean approximation ratio $\Ra$ increases monotonically from $0.678$ at $\Omega_t=1\,\mathrm{rad}/\mu\mathrm{s}$ to $0.721$ at $\Omega_t=15.7$. 
For every instance it increases between $\Omega_t=2\,\mathrm{rad}/\mu\mathrm{s}$ and $15.7\,\mathrm{rad}/\mu\mathrm{s}$ (mean increase of $\Ra$ is $+0.038$, observed for all 20 instances). This increase
correlates with $\eta$ (Spearman $\rho_S=0.60$, $P=0.005$). 

For warm start with the sign $s=+1$ the dependence of $\Ra$ on  $\Omega_t$ is not monotonic ($0.851$, $0.854$, $0.844$, $0.842$, $0.861$), but
the increase from $\Omega_t=2\,\mathrm{rad}/\mu\mathrm{s}$ to $15.7\,\mathrm{rad}/\mu\mathrm{s}$ (mean increase of  $\Ra$ $+0.007$, observed for 10 of 20 instances) correlates even more
strongly with $\eta$ ($\rho_S=0.72$, $P=4\times10^{-4}$).  For well-embedded instances
the edge part of $H_C$, acting during the mixer stage, is not harmful, and a slower mixer costs little. 

In the rest of this paper we
use $\Omega_t=15.7\,\mathrm{rad}/\mu\mathrm{s}$, which is the largest value.  At this value, the
cold-start optimum at $p=1$ ($R_a=0.721$) is already close to what layer-wise optimization reaches
at $p=5$ with $\Omega_t=2\,\mathrm{rad}/\mu\mathrm{s}$ ($R_a=0.730$, obtained with the same layer-wise optimizer).

\section{Two landscape branches}
\label{sec:branches}

\subsection{Depth-one landscape}
\begin{figure*}[!t]
\centering
\includegraphics[width=0.94\textwidth]{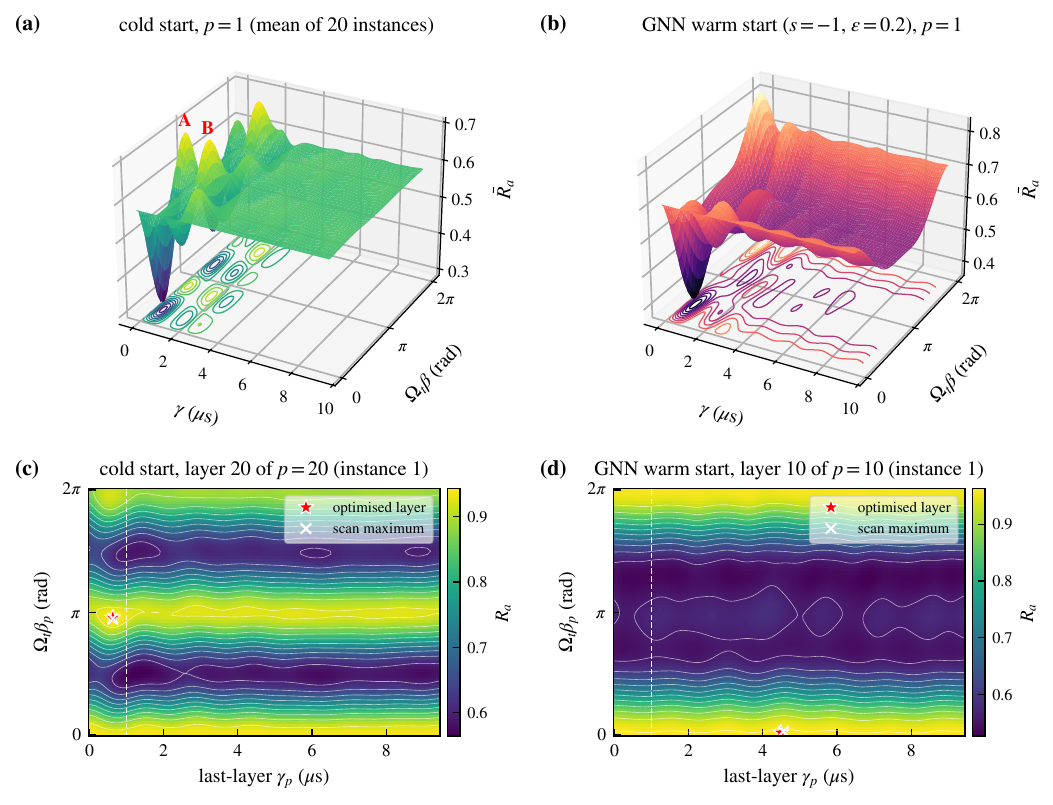}
\caption{Landscapes of the approximation ratio $R_a$ at $\Omega_t=15.7$~rad/$\mu$s, computed with the independent emulator of Sec.~\ref{sec:methods}.
(a) Depth-one landscape $\Ra(\gamma,\beta)$ for cold start averaged over the 20 instances,
with the floor showing contour lines. A and B mark two basins of local maxima; deeper circuits started from basin A form the annealing branch, and those started from basin B form the non-adiabatic branch.
(b) The same for the warm start ($s=-1$, $\varepsilon=0.2$). (c) Landscape of the last layer of
the optimized $p=20$ annealing-branch schedule of instance 1, with all other layers fixed. Star:
optimized layer; cross: maximum of the scan; dashed line: $\gamma=1\us$. (d) The same for the
last layer of the optimized warm-start schedule at $p=10$. In (c) and (d), for $\beta_p\to0$ the last cost
segment has no effect, so along the bottom edge the landscape is flat in $\gamma_p$ and equal to
$\Ra$ of the schedule without the last layer.}
\label{fig:landscape}
\end{figure*}

Figure~\ref{fig:landscape}(a) shows the cold-start $p=1$ landscape averaged over the twenty instances. As a
function of $\Omega_t\beta$, it is periodic, and has maxima near $\Omega_t\beta\approx0.75\pi$ and
$1.75\pi$, as expected for an approximately transverse rotation. As a function of $\gamma$, it
oscillates for $\gamma\lesssim3\us$, and is flat beyond this range, because the phases $\gamma(E_z-E_{z'})$
of the incommensurate spectrum of $H_C$ dephase. Two families of maxima stand out: basin A at
$\gamma\approx0.26$--$0.37\us$ and basin B at $\gamma\approx1.5$--$1.9\us$. Which one holds the
global maximum, depends on the particular instance. Basin B wins in 13 instances and basin A in 7
(instances 1, 6, 11, 12, 15, 19, 20). Where B wins, the approximation ratio is higher by up to $0.08$.

The warm-start
landscape [Fig.~\ref{fig:landscape}(b)] is different: there is a maximum at short $\gamma$, which has the highest value in 18 of  20 instances (by $0.06$ on average). In the remaining instances the long-$\gamma$
maximum wins, being ahead by at most $0.01$. For warm start, the highest maximum value almost always lies at
short $\gamma$, while for the cold start this is not the case at least for 13 instances.

From Fig.~\ref{fig:landscape}(a) alone it is not possible to conclude which basin should be used to build a deeper circuit.
A layer-by-layer optimizer simply takes the highest value, which belongs to the basin B for most instances, while the short cost segments keep the circuit in basin A. 

We considered both options. The circuits, built on basin B, form the \emph{non-adiabatic branch}
(Sec.~\ref{sec:nonadiabatic}), and the circuits, built on basin A, where all cost segments remain short, form
the \emph{annealing branch} (Sec.~\ref{sec:annealing}). The last-layer landscapes are illustrated in
Figs.~\ref{fig:landscape}(c) and \ref{fig:landscape}(d).

\subsection{The non-adiabatic branch}
\label{sec:nonadiabatic}

The non-adiabatic branch is the one which is obtained by the standard optimization procedure, without
any additional constraints. A layer-by-layer optimizer, that is good at its job, finds the global $p=1$ maximum. It means that for 13 of  20 instances it builds the deeper circuit, basing on basin B. The transfer optimizer without constraint reaches the approximation ratio $\Ra=0.766$ at $p=5$
and saturates at $0.770$ at $p=10$. 

Using twenty times more optimization steps, we observe a layer-wise baseline, which reaches $0.837$ at $p=5$, still gaining about $0.01$ per layer. Seeding its pool with the TQA schedule does not change this value ($R_a=0.840$), because the block search of the first layer always returns the parameters from basin B, whatever the seed. Therefore, both optimizers end up in basin B, and $\Ra$ grows slowly,
or saturates with depth $p$. This is the reason why we introduce the additional constraint on $\gamma_k$ in route A.

Two basins differ in phase, which is imprinted by one cost segment. The mixer couples bitstrings
that differ by one spin flip. Between two such strings, the cost segment produces the relative
phase $\gamma\Delta E$, where $\Delta E$ is the change of $H_C$ under the flip. Averaged over all
strings and atoms, $|\Delta E|=1.8$--$2.3$~rad/$\mu$s for all studied instances. In basin A this phase is
$\gamma|\Delta E|=0.5$--$0.7$~rad, comparable with the effective transverse rotation of the same
layer, $\theta^{\rm eff}=0.55$--$0.71$~rad (Sec.~\ref{sec:annealing}). 

This layer is a small step
of a discretized anneal from the transverse field to $H_C$, and the state of the system can follow the ground
state of the layer generator. In basin B, the phase is in the range $3.1$--$3.7$~rad, i.e., about $\pi$. It is
around four times larger than the phase of transverse rotation, which is in range $0.71$--$1.10$~rad. 

The generator of the
first layer is then dominated by the term $\gamma H_C$ close to the end of the annealing path, while the
state is still close to $|+\rangle^{\otimes N}$. It means that the first layer is a sudden jump, rather than an
adiabatic step. It gives a better state at $p=1$, but later layers cannot correct it. This is consistent with general expectation
that QAOA improves systematically with depth only along the quasi-adiabatic
paths~\cite{Zhou2020,Wurtz2022Counter,Brady2021}. The ground-state analysis in Sec.~\ref{sec:annealing} confirms this directly [Fig.~\ref{fig:sched}(g)]: in 12 of  13 basin-B instances the first layer already leaves the state far from the ground state of the layer generator, and the later
layers do not bring it back.

\subsection{The annealing branch}
\label{sec:annealing}

\begin{figure*}[!t]
\centering
\includegraphics[width=\textwidth]{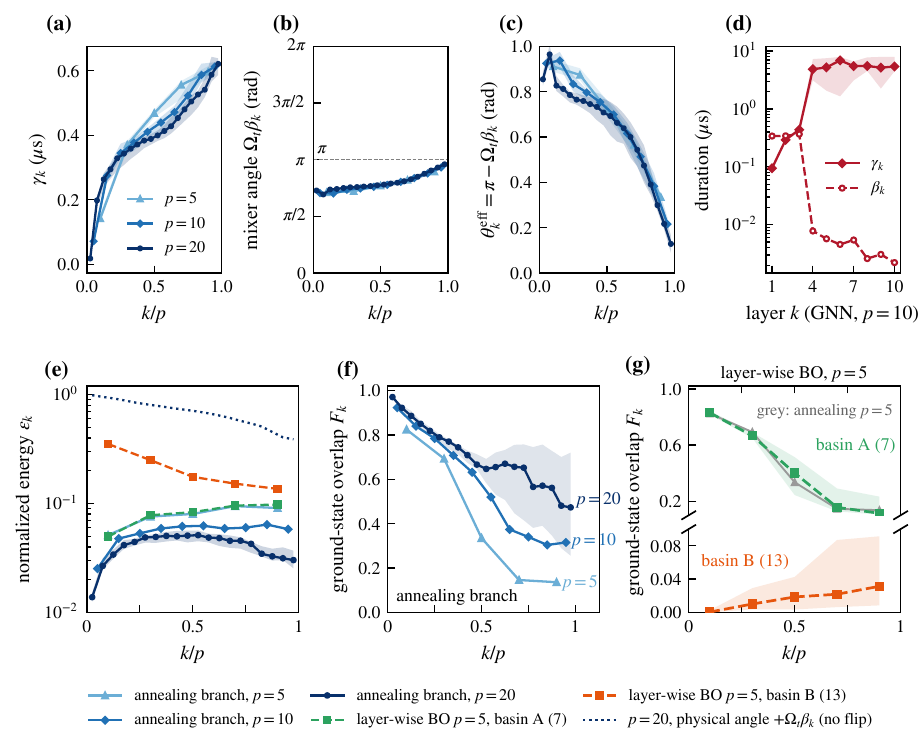}
\caption{Optimized pulse schedules [see Fig.~\ref{fig:schematic}(b)]. (a)--(c) Annealing branch, cold start. The dependences of   cost
duration $\gamma_k$ (a),   mixer angle $\Omega_t\beta_k$ (b) and   effective mixing angle
$\pi-\Omega_t\beta_k$ (c) are plotted
against $k/p$ for  $p=5$, $10$, $20$. The median over instances (lines) and interquartile range (bands) are shown. The effective mixing angle $\pi-\Omega_t\beta_k$ is equivalent to $\Omega_t\beta_k$ up to a global spin flip.
(d) GNN warm start, the median $\gamma_k$ (diamonds, interquartile band) and $\beta_k$ (open
circles) are shown as a function of layer number $k$ for $p=10$. (e) Normalized energies $\epsilon_k$ of the emulated state after layer $k$ as a function of $k/p$. Circles: The median over instances  with respect to the layer generator $H_k^{\rm eff}$ from Eq.~\eqref{eq:Heff} Band: interquartile range at $p=20$. Dots:  same states measured against the generator with the
physical angle $+\Omega_t\beta_k$. Squares: layer-wise BO at $p=5$, split by the basin of the
depth-one optimum. (f),(g) Overlap $F_k$ of the $\Pi$-even component of the state with the ground state
of $H_k^{\rm eff}$: (f) annealing branch, $p=5$, $10$, $20$ (band: interquartile range at $p=20$);
(g) layer-wise BO at $p=5$ for basin A and basin B (bands: interquartile ranges), with the
annealing branch at $p=5$ in grey, for reference. The vertical axis of (g) is broken: the upper part
covers $0.1$--$1$, the lower part $0$--$0.1$.}
\label{fig:sched}
\end{figure*}

The bound of the annealing branch is defined by the estimates from Sec.~\ref{sec:nonadiabatic}. The phase per spin flip reaches $\pi$ at $\gamma=1.4$--$1.7\us$. This is the range where basin B lies, while for $\gamma\le1\us$ it remains below $2.3$~rad in every instance. We therefore impose the constraint for the cost segments as $\gamma_k\le1\us$.

In the case when only the first cost segment is restricted to $\gamma_1\le1\us$, which selects basin A, the transfer optimizer reaches $R_a=0.854$ at $p=5$. Restricting all cost segments to $\gamma_k\le1\us$ gives
$R_a=0.872$ at $p=5$, $R_a=0.926$ at $p=10$ and $R_a=0.959$ at $p=20$, with $\Ra$ increasing with depth. At $p=5$ this is $0.035$ above the layer-wise baseline ($P=0.008$) for $5\%$ of its evaluations, and at $p=10$ it lies above the baseline's value from $p=5$  on all 20 instances. At $p=10$ the constrained and unconstrained transfer runs differ by $0.157$, with
the constrained run gives higher values in every instance.

This is explained by the optimized laser pulse schedules. Figures~\ref{fig:sched}(a) and \ref{fig:sched}(b) show how the values of $\gamma_k$ and $\Omega_t\beta_k$ of the
annealing branch are varied against $k/p$ for $p=5$, $10$ and $20$. Within the interquartile spread over
instances, the dependences for three different depths collapse onto one curve. This is the feature  that makes INTERP and
transfer effective~\cite{Zhou2020,Brandao2018}. The cost durations increase from $0.02$ to
$0.62\us$. The mixer angle $\Omega_t\beta_k$ remains between $0.70\pi$ and $0.96\pi$, and
\emph{increases} along the sequence.

At first sight, that contradicts with the decreasing mixer of a
discretized anneal. This is explained as follows. First, the prepared state
$|+\rangle^{\otimes N}$ is the \emph{highest} eigenstate of the drive term
$(\Omega_t/2)\sum_iX_i$, but not its ground state. An anneal along
$(1-\tau)(\Omega_t/2)\sum_iX_i+\tau H_C$, $\tau:0\to1$, which starts from the state $|+\rangle^{\otimes N}$, ends in the highest state of $H_C$, which is the
worst cut. Reaching the best cut from $|+\rangle^{\otimes N}$ requires a transverse field of the
opposite sign, i.e., a mixer run backwards in time. This is exactly supplied by a global spin flip. 

In the frame of the drive, the mixer segment reads as
\begin{align}
e^{-i\beta_kH_M}={}&e^{-i\theta_k\sum_iX_i/2}\nonumber\\
&\times\mathcal T\exp\Bigl[-i\!\int_0^{\beta_k}\!H_C^I(t)\,dt\Bigr],
\label{eq:dressed}
\end{align}
where $\theta_k=\Omega_t\beta_k$ is the mixer rotation angle, $\mathcal T$ denotes time ordering, and
\begin{equation}
H_C^I(t)=e^{\,i\Omega_t t\sum_iX_i/2}\,H_C\,e^{-i\Omega_t t\sum_iX_i/2}
\label{eq:HCI}
\end{equation}
is the cost Hamiltonian $H_C$ written in the frame rotating with the drive. 

Moreover,
$e^{-i\theta\sum_iX_i/2}=(-i)^N\,\Pi\,e^{\,i(\pi-\theta)\sum_iX_i/2}$ for any $\theta$. Here $\Pi=\prod_iX_i$ flips every atom, $|g\rangle\leftrightarrow|r\rangle$. Since $H_C$ contains
only the products $Z_iZ_j$ and the drive only $X_i$, $\Pi$ commutes with $H_C$ and $H_M$
($\Pi H\Pi=H$). The operator $\Pi$ for each layer can be moved toward the end of the sequence, where
it maps every measured bitstring to its complement with the same cut value. 

Up to this flip, each
layer contains a transverse rotation of the required negative sign, with the magnitude
$\theta^{\rm eff}_k=\pi-\Omega_t\beta_k$, which falls from about $0.9$ to $0.13$~rad
[Fig.~\ref{fig:sched}(c)] with increase of $\gamma_k$.

The interaction that occurs during the mixer is not
plain $H_C$: in Eq.~\eqref{eq:HCI} each $Z_i$ becomes $Z_i\cos\varphi+Y_i\sin\varphi$ with
$\varphi=\Omega_t t$. To the first order of Magnus expansion, the time-ordered factor of
Eq.~\eqref{eq:dressed} equals $\exp(-i\beta_k\bar H_C)$ with the drive-averaged interaction written as
\begin{align}
\bar H_C=\sum_{i<j}\frac{V_{ij}}{4}\Bigl[&a_k\,Z_iZ_j+b_k\,Y_iY_j\nonumber\\
&+m_k\,(Z_iY_j+Y_iZ_j)\Bigr].
\label{eq:Hbar}
\end{align}
The coefficients in Eq.~\eqref{eq:Hbar} are the averages of $\cos^2\varphi$, $\sin^2\varphi$ and $\sin\varphi\cos\varphi$ over
$\varphi\in[0,\theta_k]$,
\begin{align}
a_k&=\frac12+\frac{\sin(2\theta_k)}{4\theta_k},\qquad
b_k=\frac12-\frac{\sin(2\theta_k)}{4\theta_k},\nonumber\\
m_k&=\frac{\sin^2\theta_k}{2\theta_k},
\label{eq:abm}
\end{align}
Here $a_k\approx0.39$--$0.48$ and $b_k\approx0.52$--$0.61$ for the selected angles.
With this replacement, the decomposition reproduces every exact layer of the annealing branch with
fidelity between $0.998$ and $1.000$ (median $0.999$), while treating the interaction during the mixer stage as pure   static $H_C$ gives only $0.90$ (Appendix~\ref{app:adiabatic}).

The above decomposition assigns a generator  to each layer
\begin{equation}
H_k^{\rm eff}=-\frac{\theta^{\rm eff}_k}{2}\sum_iX_i+\beta_k\bar H_C+\gamma_kH_C ,
\label{eq:Heff}
\end{equation}
where the three terms of the equation are the flipped transverse rotation, the drive-averaged
interaction, acting during the mixer and described by Eq.~\eqref{eq:Hbar}, and the cost segment, respectively. Here $H_k^{\rm eff}$ represents
a point $k$ on the path between a transverse field with ground state  $|+\rangle^{\otimes N}$ and
$H_C$. 

We tested, whether the emulated state follows the instantaneous ground state of this path
[Fig.~\ref{fig:sched}(e)--(g)]. On the annealing branch at $p=20$, the state after the first layer has
an overlap with the ground state of $H_1^{\rm eff}$ in the range $F_1=0.95$--$0.98$  (here $F_k$ denotes the overlap
of the state after layer $k$ with the ground state of $H_k^{\rm eff}$). 

The energy of the state normalized to the spectrum of $H_k^{\rm eff}$,
$\epsilon_k=(\langle\psi_k|H_k^{\rm eff}|\psi_k\rangle-E_{\min}^{(k)})/(E_{\max}^{(k)}-E_{\min}^{(k)})$,
where $|\psi_k\rangle$ is the state after layer $k$ and $E_{\min,\max}^{(k)}$ are the extremal eigenvalues of $H_k^{\rm eff}$ ($\epsilon_k=0$ at the ground state, $1$ at the top), remains below $0.061$ at every layer of every instance. 

The largest value of $\epsilon_k$ decreases as
the schedule is refined: it reaches $0.115$, $0.078$ and $0.061$ at $p=5$, $10$ and $20$. The overlap with
the single ground state reduces to about $0.5$ at the end of the schedule, where the low-lying
states of $H_C$ become close in energy (and, in three instances, the ground state is not the target
optimum, Sec.~\ref{sec:instances}), but the energy remains at the bottom of the spectrum. Compared
against the generator with the physical angle $+\Omega_t\beta_k$ instead, the same states start at
$\epsilon_1=0.98$--$0.99$, at the top of the spectrum, which confirms that the global flip is
essential to the interpretation.

The layer-wise BO schedules, which follow the non-adiabatic branch on the basin-B instances
(Sec.~\ref{sec:nonadiabatic}), are separated by basin [Fig.~\ref{fig:sched}(g)]: in 7
instances in basin A they also remain near the ground state ($\epsilon_k\le0.11$), while in 12 of
the 13 instances in basin B the first layer already places the state at $\epsilon_1=0.26$--$0.39$,
with a ground-state overlap close to zero, and  later layers do not return it to the bottom of
the spectrum. 

Up to a global spin flip, the optimizer has therefore found a discretized annealing
path between the initial transverse field and $H_C$~\cite{Farhi2000,Kadowaki1998,Albash2018,Zhou2020,Sack2021}. The basin B is a departure from that path in the first layer. 

\subsection{Last-layer landscapes}
\label{sec:lastlayer}

Figures~\ref{fig:landscape}(c) and \ref{fig:landscape}(d) use the optimized deep schedules of
instance 1. Then we fix all layers except the last one, and scan $(\gamma_p,\beta_p)$. At $\beta_p=0$ the
last cost segment is diagonal, and it does not change the measured distribution. Therefore, the bottom edge of each of these two figures gives $\Ra$ of the schedule without its last layer. The difference between the maximum and the
bottom edge is then the contribution of the last layer.

For the annealing branch at $p=20$ [Fig.~\ref{fig:landscape}(c)], the approximation ratio reaches $\Ra=0.937$ at the bottom edge of Fig.~\ref{fig:landscape}(c).
The scan maximum gives $\Ra=0.943$ at $\gamma_{20}=0.63\us$ and $\Omega_t\beta_{20}=0.95\pi$, which coincides
with the optimized layer. 

The landscape depends mainly on $\beta$: at the optimal $\gamma$ the value
of $\Ra$ changes by $0.35$, when $\beta$ is varied. In optimal $\beta$ it changes only by $0.05$, when
$\gamma$ is varied. For $\gamma>1\us$ (right-hand side from the dashed line) the bottom-edge value is not
exceeded anywhere. Therefore, at the last layer the constraint is not active, and the unconstrained
optimization of this layer would give the same point. The last layer still is a finite
rotation and it increases $\Ra$, as expected, since $\Ra$ continues increase up to $p=20$.

The warm start at $p=10$ [Fig.~\ref{fig:landscape}(d)] behaves differently. Here, the bottom edge of the figure
gives $R_a=0.962$, and the maximum value is only $0.001$ higher ($0.963$ at $\Omega_t\beta_{10}=0.025\pi$,
$\gamma_{10}\approx4.6\us$, which again coincides with the optimized layer). At such a small
rotation the landscape hardly depends on $\gamma$ (with variation $0.005$). The value of
$\gamma_{10}$ does not matter, while a rotation with $\Omega_t\beta\approx\pi$ reduces $\Ra$ to
$0.56$--$0.59$. The last layer therefore has almost no effect.

After nine layers, the optimal cut
of this instance is already measured with probability $0.77$, and a large rotation can only spread
the state over other bitstrings. For comparison, in the $p=3$ warm-start schedule for the same
instance, the last layer increases $\Ra$ from $0.902$ to $0.951$, with a short cost segment
$\gamma_3=0.58\us$. The warm start thus uses short cost segments without any constraint, in
agreement with Fig.~\ref{fig:landscape}(b). After a few layers, more steps bring almost
nothing; we return to these long layers in Sec.~\ref{sec:warm}.

In the landscapes of Fig.~\ref{fig:landscape},  two cold-start branches separate already
at the first layer: the annealing branch starts in basin A and remains there at short cost segments up
to the last layer [Fig.~\ref{fig:landscape}(c)], while the non-adiabatic branch starts in basin
B and never comes back. For warm start, this separation is not observed, since its $p=1$ maximum is
already found at short $\gamma$ [Fig.~\ref{fig:landscape}(b)]. 

For optimization of the analog sequences,
this means that the greedy choice at small depth can be wrong. A physically motivated
constraint is cheaper than a broader search: the constrained run used the same time budget as the
unconstrained one.

\begin{figure*}[!t]
\centering
\includegraphics[width=\textwidth]{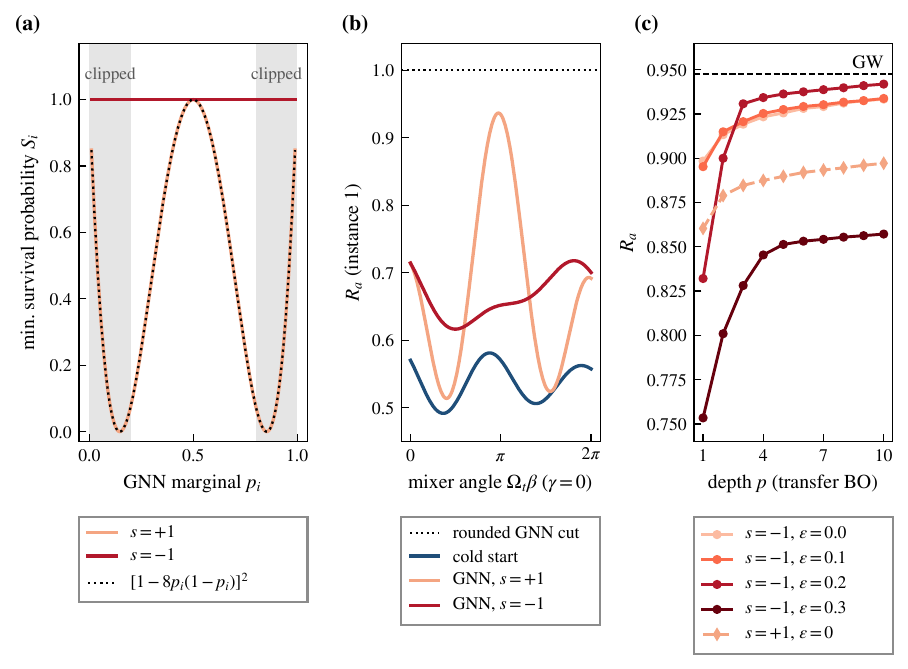}
\caption{Effects of warm start and choice of the mixer sign. (a) Smallest  survival probability of the
single-atom prepared state during the mixer stage, $S_i=\min_t|\langle\psi_i|e^{-iH_{M,i}t}|\psi_i\rangle|^2$,
Eq.~\eqref{eq:survival}, for $s=\pm1$ (single atom, $H_C$ omitted).
Gray shading: marginals $p_i<\varepsilon$ and $p_i>1-\varepsilon$ for $\varepsilon=0.2$; clipping moves them to $p_i=\varepsilon$ or $1-\varepsilon$, and after clipping there are no single-atom states in the shaded regions. Without clipping, $78\%$ of the GNN marginals of our instances fall in the shaded regions, i.e., have $p_i<0.2$ or $p_i>0.8$. (b) Many-body evolution under a single
mixer segment ($\gamma=0$) for instance 1 with $\varepsilon=0.2$. Dotted line: cut ratio of the
rounded GNN output. (c) $\Ra$ versus depth for the transfer optimizer with $s=-1$ and four
clipping values, and with $s=+1$ ($\varepsilon=0$).}
\label{fig:warm}
\end{figure*}
\section{Warm start and mixer sign}
\label{sec:warm}

Figure~\ref{fig:warm}(a) shows the single-atom survival probability of the initially prepared state $|\psi_i\rangle$, which is defined in Eq.~\eqref{eq:survival}, as a function of the GNN marginal $p_i$. For
$s=-1$ the survival probability is one for every $p_i$, since the initially prepared state $|\psi_i\rangle$ is an eigenstate
of the mixer. 

With $s=+1$ every atom with $p_i\ne1/2$ is rotated away from its GNN prepared state
during the mixer stage, and the atoms with $p_i\approx0.15$ or $p_i\approx0.85$ are rotated to the orthogonal state.
Such values are typical for  the network output: $78\%$ of the 240 marginals of our instances lie
in the range $[0.0,0.2]\cup [0.8,1.0]$, which is shown  as shaded region in Fig.~\ref{fig:warm}(a). About half of these values lie in the range within $0.05$ of the two zeros of
$S_i$.

For $s=+1$ and $\varepsilon=0$, more than half of the atoms ($54\%$) therefore have
$S_i<0.1$. The clipping of $p_i$ regions does not cure this. It moves the marginals from shaded regions to
$p_i=0.2$ or $0.8$, while $S_i=0.08$ remains small, and the mean survival probability changes only from
$0.20$ to $0.18$. Whether the warm state is preserved is not defined by clipping, but by the sign $s$. 

For $s=-1$ the clipping plays a different role: it leaves $S_i=1$, unchanged but keeps the mixer Rabi frequency $\Omega_i=2\Omega_t\sqrt{p_i(1-p_i)}$ from Eq.~\eqref{eq:mix} above
$0.8\,\Omega_t$. Therefore, no atom is frozen in its prepared state.

For many-body system the difference between  two mixer signs is equally clear. Figure~\ref{fig:warm}(b) shows the dependence of $\Ra$ on the mixer angle $\Omega_t\beta$ for instance 1 after the preparation,  and after a single
mixer segment with $\gamma=0$. 

With $s=-1$, the approximation ratio $\Ra$ changes only through $H_C$ terms, acting
during the mixer, and remains in the range between $0.62$ and $0.72$. With $s=+1$ it swings between $0.51$ and
$0.94$. Near $\Omega_t\beta=\pi$ the mixer reflects each Bloch vector along the field axis. This sharpens the marginals toward the rounded GNN string or toward its global flip (the rounded cut is optimal for this instance). The $s=+1$ mixer, therefore, acts part as a rounding operation.
That helps at $p=1$ but leaves little room for later layers to improve.

The depth dependence of $R_a$, shown in  Fig.~\ref{fig:warm}(c), confirms this. With $\varepsilon=0$, the
transfer optimizer gives $0.899$ at $p=1$ and $0.933$ at $p=10$ for $s=-1$, against $0.860$ and
$0.897$ for $s=+1$. With the layer-wise baseline the picture is the same: at $p=5$, $s=-1$
with $\varepsilon=0.2$ reaches $0.940$ against $0.895$ for $s=+1$ with $\varepsilon=0$
(18 of 20 instances, $P=3\times10^{-4}$), although $s=+1$ is ahead at $p=1$ ($0.859$ versus
$0.833$). 

The clipping parameter trades the head start against the ability to move, since the mixer acts weakly on atoms with $p_i$ close both to 0 or 1 [Fig.~\ref{fig:warm}(a), shaded regions]. At
$\varepsilon=0$ and $0.1$ the curves start high and flatten near $0.933$; at $\varepsilon=0.2$
the first layer gives only $0.832$, but $\Ra$ climbs to $0.931$ at $p=3$ and $0.942$ at $p=10$.
At $\varepsilon=0.3$ performance drops to $0.857$ at $p=10$. 

We have not analyzed this regime in
detail; with $2\sqrt{p_i(1-p_i)}\ge0.91$ the warm start is close to the cold start, and we
suspect that, without the $\gamma$ constraint, the optimizer then meets the same non-adiabatic
trap as the cold start. The existence of an intermediate optimum in $\varepsilon$ is consistent
with the role of regularization in warm-started QAOA~\cite{Egger2021,Tate2023Quantum,He2026RWS}
and with the observation that QAOA started too close to a classical string tends to stay
there~\cite{Cain2022}.

Figure~\ref{fig:sched}(d) shows how the warm-started optimizer uses depth. The first three
layers are short, with $\gamma_k\lesssim0.45\us$ and large mixer angles. From the fourth layer
on, the optimizer appends layers with $\gamma_k\approx5\us$ and $\beta_k\approx0.005\us$: long
free evolution followed by a small rotation. These layers raise $\Ra$ by about $0.01$ between
$p=3$ and $p=10$, but the total pulse time grows from $2.0$ to $38\us$ (Sec.~\ref{sec:sampling}).

The last-layer landscape of the $p=10$ schedule [Fig.~\ref{fig:landscape}(d),
Sec.~\ref{sec:lastlayer}] shows the reason: the optimized last layer has $\beta_p\to0$, where the
final cost segment is diagonal and has no effect on any measured quantity. The value $\gamma_p\approx5\us$
is arbitrary and only increases the pulse time. For hardware purposes, the relevant warm-start result is the one at $p=3$.

\begin{figure*}[!t]
\centering
\includegraphics[width=\textwidth]{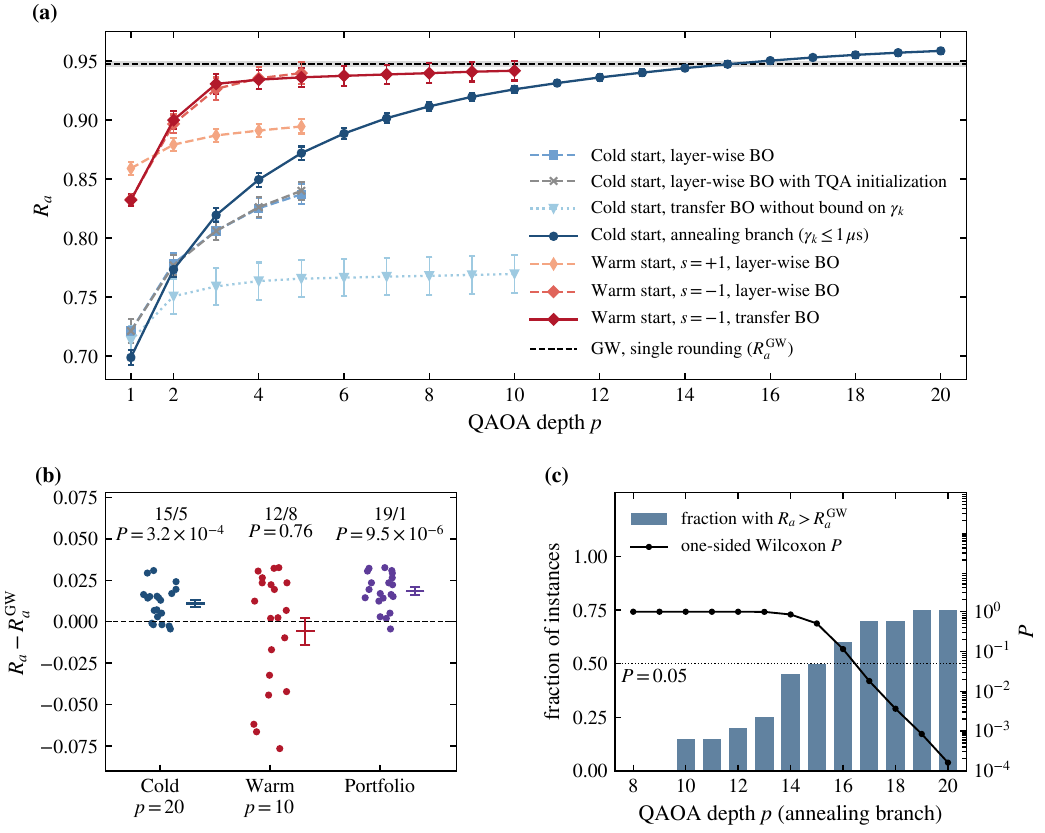}
\caption{Summary of simulation results at $\Omega_t=15.7$~rad/$\mu$s. (a) Approximation ratio $\Ra$ versus QAOA
depth $p$ for all protocols, averaged over the 20 instances; error bars are standard errors. Dashed line and gray band represent mean and standard error of the single-rounding GW
expectation $\RGW$, respectively, in the same instances. (b) Per-instance difference $\Ra-\RGW$ for 
annealing branch at $p=20$ (cold), warm start ($s=-1$, $\varepsilon=0.2$, transfer BO) at
$p=10$ (mean of three repetitions; warm) and their portfolio, defined $\max(\Ra^{\rm cold},\Ra^{\rm warm})$.
Dots are individual instances, bars are the mean value and its standard error; the numbers give wins/losses
and the two-sided Wilcoxon $P$. (c) Annealing branch: fraction of the instances with $\Ra>\RGW$
(bars, left axis) and one-sided Wilcoxon $P$ for $\Ra>\RGW$ (line, right axis); the dotted
line marks $P=0.05$ level.}
\label{fig:main}
\end{figure*}
\section{Comparison with Goemans--Williamson}
\label{sec:gw}

Figure~\ref{fig:main}(a) summarizes the comparison of the QAOA schemes, which we considered, with the classical GW algorithm. The cold start annealing branch crosses the GW
expectation value between $p=15$ and $p=16$. At $p=15$ the mean difference between them is $-0.0003$ (10 wins, 10
losses). The one-sided test for $\Ra>\RGW$ falls below $0.05$ at $p=17$ ($P=0.018$) and reaches
$2\times10^{-4}$ at $p=20$, where the mean difference is $+0.011\pm0.002$ and 15 of 20 instances
lie above GW [Fig.~\ref{fig:main}(b,c)]. 

The warm start is statistically indistinguishable from
GW: $-0.017$ at $p=3$ ($P=0.15$) and $-0.006$ at $p=10$ ($P=0.76$, 12 wins), with a much larger
spread between the instances [Fig.~\ref{fig:main}(b)].

\subsection{Origin of the spread between the instances}

\begin{figure*}[!t]
\centering
\includegraphics[width=\textwidth]{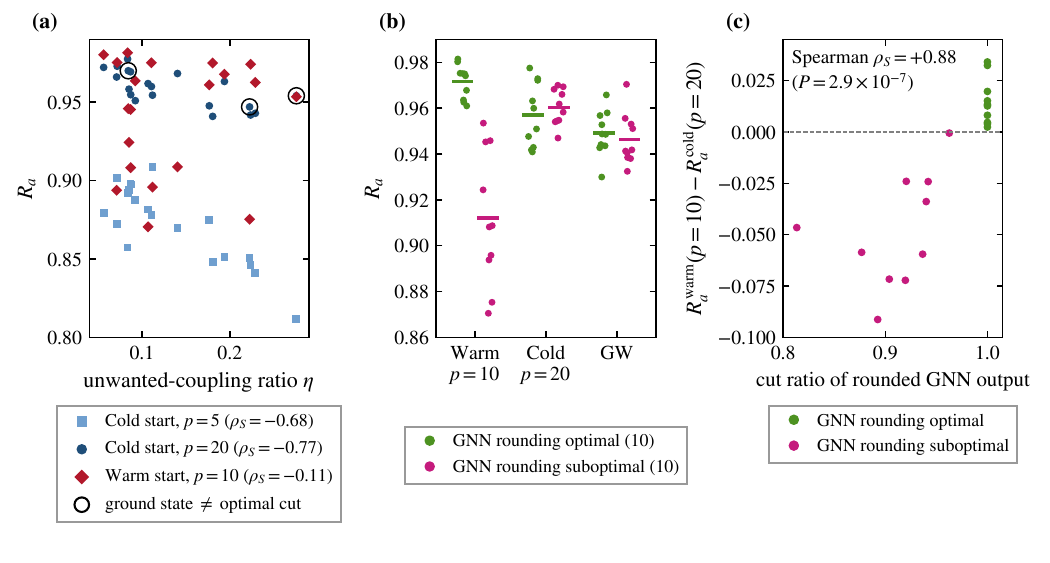}
\caption{Per-instance performance. Cold start: annealing branch
($\gamma_k\le1\us$, transfer BO); warm start: $s=-1$, $\varepsilon=0.2$, transfer BO (mean of three
repetitions). (a) $\Ra$ as a function of the unwanted-coupling ratio $\eta$, Eq.~\eqref{eq:eta}, for cold
start at $p=5$ and $p=20$ and warm start at $p=10$; Spearman coefficients $\rho_S$ are listed in the
legend. Open circles mark three instances for which the ground state of Eq.~\eqref{eq:HC} is not an
optimal cut. (b) $\Ra$ of warm start ($p=10$), cold start ($p=20$) and the GW
single-rounding expectation, separately for 10 instances in which the rounded GNN output is
an optimal cut (green) and for 10 instances, in which it is not (magenta). The horizontal bars are group mean values.
(c) The difference between warm start at $p=10$ and cold start at $p=20$ as a function of the cut ratio of
the rounded GNN output, colored similarly to (b). Dashed line is zero level.}
\label{fig:embed}
\end{figure*}

Figure~\ref{fig:embed} shows where this spread comes from. Two optimization routes depend on different
properties of an every instance:  the annealing branch depends on the quality of the embedding
[Fig.~\ref{fig:embed}(a)], and the warm start depends on the quality of the network output
[Fig.~\ref{fig:embed}(b),(c)]. 

For warm start, we split the instances by whether the rounded GNN output is an optimal cut or not [Fig.~\ref{fig:embed}(b)]. In 10 instances where it is an optimal cut, the warm start reaches $R_a=0.972$ at $p=10$, well above both the GW ($0.949$) and the annealing branch ($0.957$). In the other 10 instances  it reaches $R_a=0.912$ (Mann--Whitney $P=2\times10^{-4}$ between groups). The annealing branch
does not show any differences ($0.957$ versus $0.960$, $P=0.62$), and neither does the GW. 

In all instances, the advantage of  warm start over the annealing branch is correlated with the cut ratio of the
rounded GNN output ($\rho_S=0.88$, $P=3\times10^{-7}$). It is positive in all 10 instances with 
optimal rounding and negative in all 10 other instances [Fig.~\ref{fig:embed}(c)]. In contrast to GNN, the annealing branch is sensitive to
the quality of the embedding. Its $\Ra$ at $p=20$ has negative correlation with the value of $\eta$
($\rho_S=-0.77$, $P=7\times10^{-5}$), while the warm start shows no such dependence
($\rho_S=-0.11$) [Fig.~\ref{fig:embed}(a)].

\subsection{Portfolio of  two routes}

From the above discussion, we see that two routes fail on different instances (Fig.~\ref{fig:embed}). Therefore, running them both and keeping
the larger expected cut is better, than either of them [right column of Fig.~\ref{fig:main}(b)]. This allows creation of the portfolio, which does not require prior knowledge, which route is better in
advance. Instead, only the estimation of both expectation values is needed. 

With  warm start at $p=10$ and the
annealing branch at $p=20$, this portfolio gives $\Ra=0.966$, which is above GW on 19 of 20 instances
($+0.018\pm0.002$, $P=1\times10^{-5}$). With warm start at $p=3$, i.e., $2\us$ pulse duration, it
still gives $0.962$ (18 of 20, $P=2\times10^{-5}$). Appendix~\ref{app:table} lists all
per-instance values.

The rounded network output, which costs one forward pass, is a second classical reference, and in
expectation it is demanding. Its mean cut ratio, $0.955$, lies above the GW expectation and
above the warm start at $p=10$ (by $0.014\pm0.007$, $P=0.006$), and neither the annealing branch
at $p=20$ ($+0.003\pm0.013$) nor the portfolio ($+0.011\pm0.011$, $P=0.45$) is significantly above
it. 

This mean value hides two groups. In 10 instances, where the rounding is optimal, no method can
exceed it. In the other 10 instances it gives $R_a=0.911$, while the annealing branch,  not using the
network, reaches $R_a=0.960$, and GW gives $R_a=0.946$. 

A quantum route, therefore, adds to the network only in the case, when the network is wrong. The route, which has this effect,  is the annealing branch, but not the warm start ($0.912$ on the same instances). The comparison of expectation values also understates the
sampled protocols: the rounding returns one string, while the quantum state is measured
repeatedly.

With $N_s=2$ samples, the expected best cut of the annealing branch at $p=20$ is
$0.988$. That exceeds the rounding on each of 10 instances, where the rounding is suboptimal.
With $N_s=10$, it is $0.9995$ (Sec.~\ref{sec:sampling}). At equal $N_s$, the appropriate classical
comparison is the repeated GW rounding, which gives $0.978$ and $0.9993$.

One may ask whether two routes can  be merged into a single pulse sequence, that starts
from the warm state and then follows the annealing branch. Another option is a gradual change from the
custom mixer to the standard one. We tried nine such variants, six of them with the mixer sign of
Eq.~\eqref{eq:mix} (Appendix~\ref{app:fusion}, Fig.~\ref{fig:fusion} and Table~\ref{tab:fusion}).
None of them reaches the warm start from Sec.~\ref{sec:warm} at $p=10$, and none removes the
deficit in the instances where the GNN rounding is wrong. There they remain between $0.85$ and
$0.91$, below the annealing branch already at $p=10$ ($0.93$). 

Combined with the annealing branch
in a portfolio, none of them does better than the warm start used here. Once the GNN bias has
entered the state, the later layers do not leave it. It means that two routes are best run separately.
\begin{figure*}[!t]
\centering
\includegraphics[width=\textwidth]{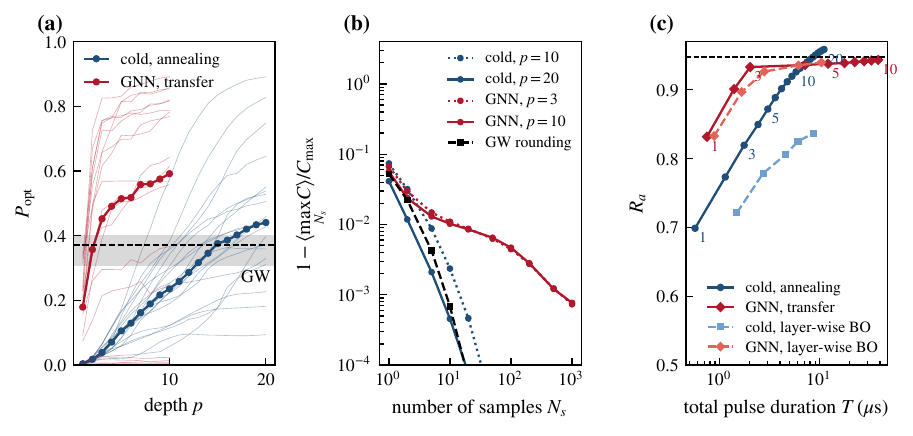}
\caption{Sampling statistics and pulse time. (a) Probability of measuring an optimal cut as a function of
depth. Median values are shown as markers, and individual instances as thin lines. The gray band is the interquartile range of GW, dashed line represents its median value. (b) Expected shortfall of the best of $N_s$ samples, 
$1-\langle\max_{N_s}C\rangle/C_{\max}$, averaged over the instances. (c) $\Ra$ as a function of the total pulse
duration $T$. The numbers indicate depth. Dashed black line is the mean value of $\RGW$. In all panels, GNN denotes  warm start with $s=-1$ (transfer optimizer unless stated otherwise).}
\label{fig:sampling}
\end{figure*}

Below we compare a sample from each method. GW is usually sampled repeatedly. At $n=12$, ten rounds already find the optimum with the probability close to one
($\langle\max_{10}C\rangle^{\rm GW}/C_{\max}=0.999$). Our comparison shows that 
analog protocols prepare states with the average cut better than that of the GW distribution. 
However, here we do not claim that these instances are solved faster than GW, or than the exhaustive search, which
is trivial at this size. The sampling comparison is the subject of the following section.

\section{Sampling and pulse duration}
\label{sec:sampling}

Figure~\ref{fig:sampling}(a) shows the dependence of the probability  of the optimal cut $\Pop$, defined in Eq.~\eqref{eq:popt}, on depth $p$. In the annealing branch it increases steadily, from a median of $0.004$ at $p=1$ to $0.24$ at $p=10$ and $0.44$ at $p=20$. It exceeds the GW
median ($0.37$) from $p\approx15$. The warm
start produces large $\Pop$ quickly (median $0.45$ at $p=3$, $0.59$ at $p=10$), but splits the
instances in two groups: in six instances $\Pop$ remains below $0.03$ at $p=10$, because the state
is concentrated near a suboptimal string. This is a bias which a warm start
produces~\cite{Cain2022}. It shows up directly in the dependence of the expected shortfall on the number of samples $N_s$, illustrated in Fig.~\ref{fig:sampling}(b). 

For the annealing branch, at $p=20$ the expected shortfall
of the best sample drops below $10^{-3}$ at $N_s=10$, similarly to the GW. For  warm start it levels off
near $10^{-2}$, and is still as low as $7\times10^{-4}$ at $N_s=1000$. In a sampling setting, the cold
annealing branch is a more reliable route, and  warm start is attractive mainly due to short sequences and higher expectation values.

Figure~\ref{fig:sampling}(c) illustrates dependence on the pulse time. The annealing branch needs about $0.55\us$ per
layer, which means $T=5.8\us$ at $p=10$ and $11\us$ at $p=20$. The warm start reaches $\Ra=0.93$ at
$T=2.0\us$ ($p=3$) and gains little afterward, while $T$ increases to $38\us$ at $p=10$ due to the
reason, explained in Sec.~\ref{sec:warm}. 

For  cold start the layer-wise baseline lies well below the annealing
branch at equal $T$; for  warm start two optimizers trace similar curves. These durations set the
cost of each route under noise (Sec.~\ref{sec:noise}).

\section{Optimization cost}
\label{sec:cost}

\begin{figure*}[t]
\centering
\includegraphics[width=0.95\textwidth]{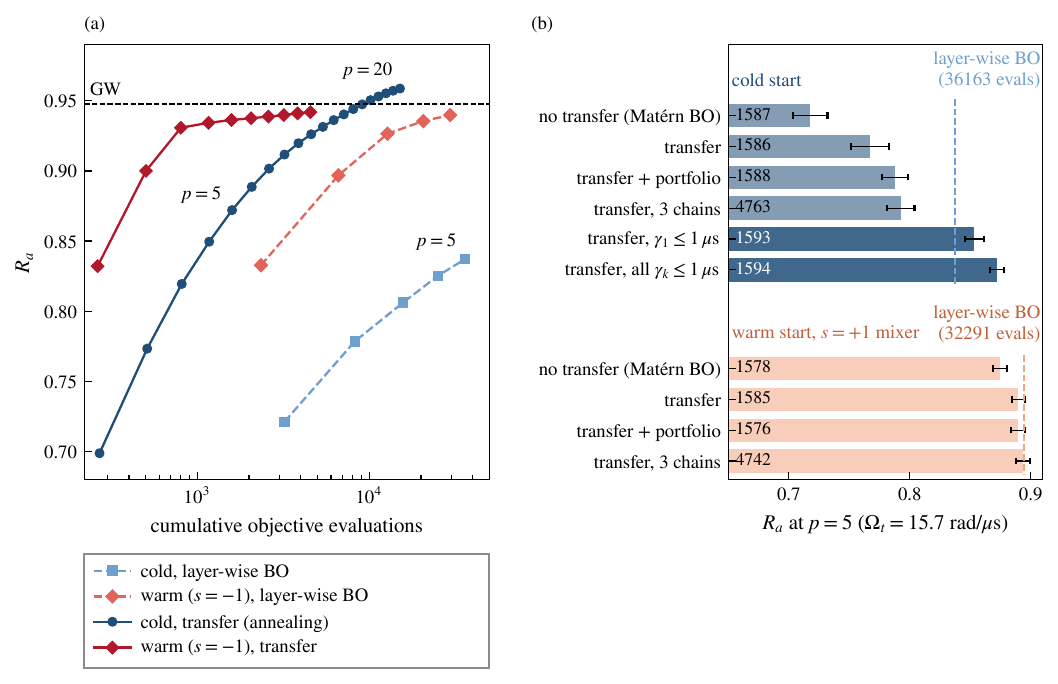}
\caption{Optimization cost. (a) The approximation ratio $\Ra$ as a function of the  cumulative number of objective evaluations per
instance (mean over instances) for the layer-wise baseline and the transfer optimizer. The
source database is not included (Sec.~\ref{sec:methods}). (b) Ablation at $p=5$: the calculated $\Ra$ for different
variants of the transfer optimizer (indicated as bars; the numbers on the bars are the evaluations per instance) and for the 
layer-wise baseline (dashed lines), both for  cold start and for warm start with $s=+1$.}
\label{fig:cost}
\end{figure*}

Figure~\ref{fig:cost}(a) plots $\Ra$ against the cumulative number of evaluations. To reach $p=5$,
the layer-wise baseline uses $36\,163$ evaluations per instance for cold start and $29\,619$
for  warm start. The transfer optimizer uses about $1\,590$, which is $23$ and $19$ times fewer, respectively.
For warm start, both give the same $\Ra$ within $0.004$ ($P=0.06$). For cold start the annealing-branch transfer is better by $0.035$. The full annealing branch up to $p=20$ costs
$15\,101$ evaluations, less than half of the baseline's cost for $p=5$.

The ablation in Fig.~\ref{fig:cost}(b) separates the ingredients at $p=5$. For  cold start,
a 30-evaluation Matérn search without transfer gives $0.718$; transferred starting points raise
this to $0.767$. The portfolio of start points, chosen to cover different source optima, or three
chains with different starts at three times the budget, give $0.788$ and $0.793$. 

All of these values remain below the layer-wise baseline ($R_a=0.837$), because they are all built on  basin B. The constraint
$\gamma_1\le1\us$ alone gives $0.854$, and the constraint, applied on every layer, gives $0.872$, both at the
budget of the plain transfer run. For  warm start, where there are no branching problems, the transfer
alone recovers the baseline to within $0.005$, and three chains to within $0.001$.

\section{Toward an experiment: control requirements and noise}
\label{sec:noise}

All results reported above were obtained for a closed unitary system evolution with perfect control. However, the experimental implementation is never noiseless. In this section we discuss the requirements for the experimental implementation and evaluate the optimized schedules for the error model of an existing  analog device, represented by an Aquilla neutral-atom quantum processor.

\subsection{Control requirements for experiment}

The cold-start protocol needs a global drive with Rabi frequency $\Omega_t=2\pi\times2.5$~MHz and
site-resolved, static detunings $\Delta_i$ of up to $7.2$~rad/$\mu$s ($2\pi\times1.15$~MHz), as defined in 
Eq.~\eqref{eq:comp}.
Global laser drives of this strength and local detuning patterns are available for current
analog devices~\cite{Wurtz2023Aquila,deOliveira2025}. The site-resolved light shifts have been already
used to encode weights~\cite{deOliveira2025,Goswami2024}. 

The warm start  also requires
site-resolved Rabi frequencies during the preparation and  mixer stages, as written in Eqs.~\eqref{eq:prep} and
\eqref{eq:mix}. The detuning pattern must be switched between the cost and mixer segments. Both
protocols need a three-dimensional atomic array with positions controlled to a fraction of a
micron~\cite{Barredo2018,Kumar2018}, since a small displacement $\delta r_{ij}$ substantially changes the
coupling energy by $\delta V_{ij}/V_{ij}=-6\,\delta r_{ij}/r_{ij}$.

\subsection{Error model and emulation}
\label{sec:noisemodel}

In the experiments with arrays of ultracold neutral atoms~\cite{deLeseleuc2018,Levine2018, Radnaev2025, Henriet2020, Evered2023, Wu2021a}, the  dominant error sources are (a) shot-to-shot  laser intensity fluctuations and spatial inhomogeneity of the Rabi frequency; (b) detuning offsets from laser frequency noise and Doppler shifts; (c) atom-position fluctuations, which
enter to the model both through variation of $V_{ij}$ and the local detunings $\Delta_i$, as written in Eq.~\eqref{eq:comp}; (d) dephasing
and spontaneous decay of the Rydberg states~\cite{Beterov2009}; and (e) detection errors. Generic arguments suggest that noise erodes the advantage of depth~\cite{StilckFranca2021,Wang2021NIBP} and changes the
QAOA landscape~\cite{Xue2021,Marshall2020}. It means that the deep annealing branch is the most at
risk result.

We use the error model which is provided with the 
Bloqade~\cite{BloqadeNoisy} package. It represents the parameters of the Aquila analog quantum processor~\cite{Wurtz2023Aquila}. All error rates in the noise model can be scaled by a
common factor $\lambda$, with $\lambda=1$ is the exact Aquila model. 

The error model includes the following parts~\cite{BloqadeNoisy}:

(i) Quasi-static errors,  which are drawn once per shot and remain constant during the laser pulse sequence: a relative Rabi
error $\delta\Omega/\Omega$ with a global part of standard deviation $0.008\lambda$ and an
independent per-atom part of $0.018\lambda$; a detuning offset with a global part of
$0.18\lambda$~rad/$\mu$s and a per-atom part of $0.50\lambda$~rad/$\mu$s; and a displacement of
every atom by $0.05\lambda\,\mu$m (standard deviation per Cartesian coordinate).  The interactions
$V_{ij}$ follow the displaced positions, while the compensation detuning of Eq.~\eqref{eq:comp}
is programmed for the nominal ones. It means that the position errors also result in the uncompensated single-site fields. 

(ii) The Markovian errors, described by a Lindblad master equation with jump operators
$\sqrt{\gamma_\phi}\,Z_i$ (dephasing, $\gamma_\phi=0.02\lambda\,\mu\mathrm{s}^{-1}$) and
$\sqrt{\gamma_1}\,|g\rangle\langle r|_i$ (decay, $\gamma_1=0.01\lambda\,\mu\mathrm{s}^{-1}$).

(iii) Detection errors: a ground-state atom is interpreted as Rydberg atom with probability $0.01\lambda$
and a Rydberg atom is measured as ground-state atom, with probability $0.08\lambda$, independently for each atom in the array. This model was specified for two-dimensional arrays. We applied it for the three-dimensional arrays as an estimate.

We solve the master equation by quantum trajectories~\cite{Dalibard1992,Molmer1993,Plenio1998},
with a matrix-free state-vector solver, written specifically for this purpose, and verified the results using Bloqade package, which has a
dense Lindblad integration and exact enumeration (Appendix~\ref{app:noise}). 

The detection errors affect the final bit-string distribution, and are included exactly in the model. We evaluate the schedules,
optimized in the noiseless environment, (first repetition of the transfer runs, no re-optimization)
at $p\in\{1,2,3,5,10,15,20\}$ for the annealing branch and $p\in\{1,2,3,5,10\}$ for  warm
start ($s=-1$, $\varepsilon=0.2$), on all 20 instances, for $\lambda\in\{0.1,0.2,0.3,0.5,1,2\}$,
with 300 trajectories per instance, depth and $\lambda$.

The quasi-static draws are shared by
all depths and both routes on an instance (common random numbers), which makes the paired
comparisons below insensitive to the sampling noise of the trajectories. Without noise, the solver
reproduces the stored values of all evaluated schedules to $2\times10^{-10}$.

Because $C$ is a sum over edges, $\langle C\rangle$ depends only on the single-atom and
pair marginals of the measured distribution, and the effect of independent detection errors on
them can be inverted exactly~\cite{Bravyi2021Readout}. We therefore quote two values:
the measured $\Ra$, which includes detection errors, and the readout-corrected $\Ra$, which is
the value for ideal detection. Unless stated otherwise $\Ra$ under noise means the measured value.

\subsection{Optimal depth under noise}
\label{sec:noisedepth}

\begin{figure*}[t]
\centering
\includegraphics[width=\textwidth]{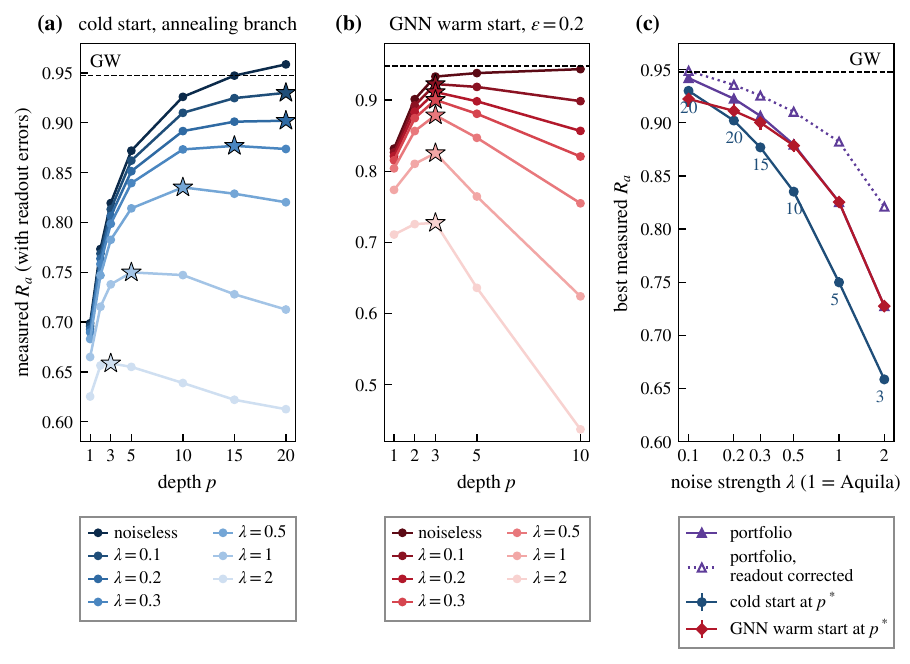}
\caption{Noisy emulation with the Aquila error model with the noise level scaled by a factor $\lambda$. (a) The measured $\Ra$  (including detection errors) as a function of depth $p$  of the annealing branch  for the noiseless case and for $\lambda=0.1$--$2$. The stars mark the best depth $p^*$ at each $\lambda$. (b) The same for GNN
warm start ($s=-1$, $\varepsilon=0.2$, transfer optimizer). (c) Measured $\Ra$ at the best depth
versus $\lambda$ for both routes (numbers: $p^*$ of the annealing branch; the warm start has $p^*=3$ at every $\lambda$), the portfolio of the two (better measured
value per instance), and the portfolio after exact readout correction. Dashed line: mean GW
expectation. Error bars: standard errors over the 20 instances; the trajectory error of each
point is below $0.002$.}
\label{fig:noise}
\end{figure*}

Figure~\ref{fig:noise}(a) shows the result of simulation in noisy environment. With the Aquila model ($\lambda=1$) the  annealing branch no longer improves with depth beyond $p=5$. The measured $\Ra$ is $0.665$,
$0.738$, $0.750$, $0.747$, $0.728$ and $0.713$ at $p=1,3,5,10,15,20$ (total durations $T=0.57$,
$1.76$, $3.0$, $5.8$, $8.5$ and $11.0\us$). The noiseless gain of $0.087$ between $p=5$ and
$p=20$ is more than canceled by an additional loss of $0.124$. The best depth $p^*$ of the
annealing branch decreases steadily with increase of the error level: $p^*=20$ at $\lambda\le0.2$, $15$ at
$\lambda=0.3$, $10$ at $\lambda=0.5$, $5$ at $\lambda=1$ and $3$ at $\lambda=2$.

The warm start behaves differently [Fig.~\ref{fig:noise}(b)]. Its best depth is $p^*=3$ for every
$\lambda$, because the gain beyond $p=3$ is small in the noiseless case ($0.933\to0.944$ for
$p=10$). The long phase layers of Sec.~\ref{sec:warm} raise $T$ from $2.0\us$ at $p=3$ to
$11.8\us$ at $p=5$ and $38\us$ at $p=10$. At $\lambda=1$ the warm start reaches a measured value
$\Ra=0.826$ at $p=3$ ($0.882$ after readout correction), against $0.750$ for the best cold-start
depth; it is better in all 20 instances ($P=1.9\times10^{-6}$). 

Figure~\ref{fig:noise}(c)
traces the comparison of the two routes at their best depths: the warm start is better on 16
instances and worse on 3 at $\lambda=0.5$ ($P=1.3\times10^{-4}$) and on 13 of 20 at $\lambda=0.3$ ($P=0.012$),
the two are indistinguishable at $\lambda=0.2$ ($0.911$ versus $0.902$, $P=0.39$), and only at
$\lambda=0.1$ does the annealing branch at $p=20$ lie above it on average ($0.930$ versus
$0.922$, 9 of 20 instances, $P=0.33$). The noiseless ordering of Sec.~\ref{sec:gw}, in which the
deep annealing branch is the strongest single route, is therefore restored only for the error
level about an order lower that of the existing Aquila model.

\begin{figure*}[t]
\centering
\includegraphics[width=\textwidth]{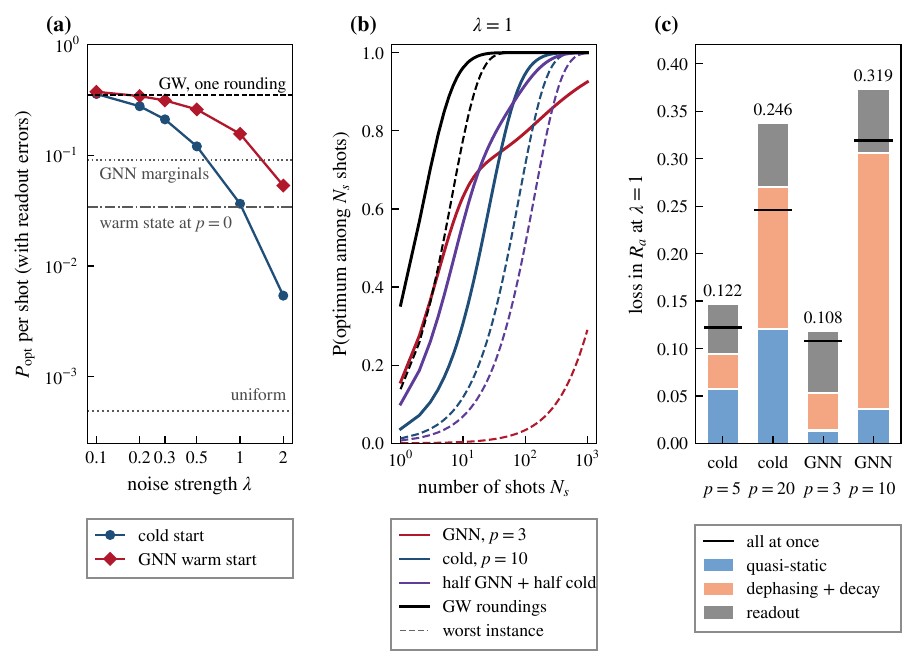}
\caption{Sampling and error budget under noise. (a) Probability $\Pop$ of measuring an optimal
cut (with detection errors), mean value over all the instances at the depth, that maximizes it, as a function of noise strength
$\lambda$. Horizontal lines from up to bottom illustrate the levels: one GW rounding, sampling the GNN marginals $p_i$ directly, the
clipped warm state before the first layer ($p=0$, $\varepsilon=0.2$), and uniform random
strings. (b) Probability of finding an optimal cut among $N_s$ shots at $\lambda=1$, mean value over all
instances (solid) and the worst instance (dashed), for  warm start ($p=3$), the annealing branch
($p=10$), half of the shots on each, and $N_s$ GW roundings. (c) Loss of measured $\Ra$, compared to the noiseless case at $\lambda=1$, with only one error class switched on (bars, stacked for
display; the errors are not additive) and with all of them on at once (black lines and numbers).}
\label{fig:noisesampling}
\end{figure*}

The same holds for the comparison with GW. With detection errors, the best protocol remains below
$\RGW$ in every instance for $\lambda\ge0.3$, and the portfolio reaches $0.942$ at
$\lambda=0.1$ ($-0.006$ relative to GW, 7 of 20 instances above). After readout correction, the
portfolio matches GW at $\lambda=0.1$ ($0.949$ versus $0.948$, 11 of 20 instances above) and
falls below it at $\lambda\ge0.2$ (7 of 20 at $\lambda=0.2$, none at $\lambda\ge0.5$). The
portfolio itself loses its value as $\lambda$ grows: at $\lambda\ge1$  warm start is the
better member on every instance. It means that the portfolio is reduced to  warm start.

\subsection{Sampling under noise}
\label{sec:noisesampling}

In a sampling analysis, we  compare the probability $\Pop$ of an optimal string per shot with GW.
Figure~\ref{fig:noisesampling}(a) shows $\Pop$ at  depth that maximizes it. At $\lambda=1$,
the warm start at $p=3$ gives $\Pop=0.157$, which is $4.6$ times larger than the value of its own initial
state ($0.034$) and $1.7$ times larger than that of sampling the GNN marginals directly ($0.092$). However, it is still less than half of a single GW rounding ($0.354$).

The annealing branch is hit harder by the noise: its best value,
at $p=10$, is just $0.037$. Both routes reach the GW value only for noise scaling factor $\lambda\approx0.1$ (warm start at
$p=5$: $0.376$; annealing branch at $p=20$: $0.360$), while in the noiseless case the annealing
branch exceeds GW starting from $p\approx15$. (Sec.~\ref{sec:sampling}).

The averaging hides a structural difference between the cold and warm start, as shown in Fig.~\ref{fig:noisesampling}(b).
In four instances (10, 12, 14 and 17, which have the suboptimal rounded GNN output) the
warm start concentrates the system state near a wrong string, with $\Pop$ between $3\times10^{-4}$ and
$4\times10^{-3}$ at $\lambda=1$. Therefore, its success probability  saturates at the following values: $0.63$ for
$N_s=10$, $0.80$ for $N_s=100$ and $0.93$ for $N_s=1000$ on average. It reaches only $0.29$ at worst instance at $N_s=1000$.

The annealing branch has no such failure mode (its smallest $\Pop$
is $0.012$). It finds the optimum of every instance within $370$ shots with the probability $0.99$,
against a median of $120$. Spending half of the shots on each route combines  two
advantages: $0.92$ on average for $N_s=100$ and $0.999$ for the worst instance at $N_s=1000$
($99\%$ success within a median of $41$ and a maximum of $649$ shots). GW needs a median of $10$
and at most $31$ roundings for the same confidence. The portfolio idea of Sec.~\ref{sec:gw}
therefore survives noise in the sampling sense, as an allocation of shots, even where it no
longer helps to increase the expectation value.

The emulation answers the question, raised by the analysis of the noiseless results. At current error rates, the
crossing with GW, which requires $p\gtrsim16$ and $T\gtrsim9\us$ without noise, does not
survive. The best experimental choice is a warm start at $p=3$ ($T=2\us$). For the expectation
value and for small shot budgets, when the optimum is needed with high confidence, it is better when combined with a short annealing-branch circuit ($p\approx5$--$10$).

 Here we did not re-optimize the laser pulse schedules under noise, and assumed that a noise model of a two-dimensional atomic array can be applied for a three-dimensional configuration. Reaching the noiseless regime of the deep annealing branch requires all error sources to be reduced by roughly an order of magnitude.

\subsection{Error budget}
\label{sec:budget}

To understand the error budget, in the simulation we tried switching only one of the error source classes at a time, with $\lambda=1$. Results are illustrated in Figure~\ref{fig:noisesampling}(c). For the annealing branch the losses grow with depth: at $p=5$ they are $0.058$
(quasi-static), $0.037$ (dephasing and decay) and $0.053$ (detection), $0.122$ in total; at
$p=20$ they are $0.121$, $0.150$ and $0.067$, $0.246$ in total. For the warm start at $p=3$ the losses
are $0.014$, $0.039$ and $0.066$, $0.108$ in total, so detection errors dominate.

At similar duration, the warm start is less sensitive to quasi-static errors than the annealing branch
($0.014$ at $T=2.0\us$ against $0.036$ for the annealing branch at $p=3$, $T=1.8\us$). This
is consistent with the preservation of the warm state under its own mixer (Sec.~\ref{sec:gnn}):
a relative amplitude error $\delta$ tilts the mixer axis for each atom by an angle of order
$\delta$, and causes  leakage of order $\delta^2$ per mixer. 

In contrast, warm start loses more due to dephasing and decay ($0.039$ against $0.017$). For  deep warm-start circuits ($p=10$, $T=38\us$) the
dephasing and decay alone cost $0.269$, which rules out them, as we anticipated in Sec.~\ref{sec:sampling}. The losses are not exactly additive: the sum of the three single-class losses exceeds the loss, when all types of errors are present simultaneously.
\section{Discussion}
\label{sec:discussion}

All graph instances studied in our work have just 12 vertices, making it straightforward to verify the obtained solutions of weighted MaxCut problem. We compare the results with the GW algorithm, using its single-rounding expectation value.  With a handful of roundings,
it finds the optimum for every instance. Similarly,  the annealing branch of the QAOA finds it with a handful of
samples. 

The comparison shows that the  analog Rydberg-based QAOA protocol, which is limited by the always-on interatomic interactions and imperfect weight embeddings, still prepares quantum states with mean cut being better, than provided by the GW rounding distribution. 

We analyzed the required optimization depth. Our approach does not show a general advantage over every classical heuristic: in expectation values, the
rounded output of the network, used for  warm start, is as good as either of the quantum routes
(Sec.~\ref{sec:gw}). This criterion has been considered for gate-based warm starts~\cite{Tate2023Quantum,
He2026RWS}. Here it is met even by a cold-start analog protocol, where the only problem-specific
ingredients are the atomic positions and the compensating detunings.

The problem of weight embedding sets the limits that no optimizer removes. In three instances, which we considered,  the ground state of the physical Hamiltonian does not represent an optimal cut, and the annealing branch degrades with the unwanted-coupling ratio $\eta$. Larger instances will make clean three-dimensional
embeddings harder to find, which is a geometric counterpart to the overhead of gadget-based
embeddings~\cite{Nguyen2023,Cazals2025,deCorrec2025}. 

The warm start is less sensitive to the parameter $\eta$
but it inherits the quality of the network's rounding. Its successes coincide with the instances
on which the network already gives the right solution,  which is in line with the critical assessments of GNN
heuristics~\cite{Boettcher2023,Angelini2023}. 

However, the enhancement from the quantum protocols comes from the annealing branch, in the instances where the network is wrong. The portfolio is useful, because these two  failure modes are not correlated, and our attempts to merge the routes into one pulse sequence
(Appendix~\ref{app:fusion}) show that the GNN bias, once in the state, is not undone by later
annealing-like layers.  

A merger would need a mechanism that lets the protocol detect and discard a wrong warm start. It can be a recursive or measurement-based step~\cite{Bravyi2020,
Finzgar2024}, rather than a fixed schedule. The two-branch structure of the optimization landscape is a general hazard for greedy layer-wise optimization of analog sequences. The depth-one maximum becomes a wrong start, when it lies on a branch that later
layers cannot extend. Finding the right branch required a physical criterion, but not a better
optimizer. 

The schedules of the annealing branch collapse across depths and instances, which is the reason why the transfer between the instances and INTERP work so well for them. With $5\%$ of the evaluations of the layer-wise baseline, the transfer optimizer reaches higher $\Ra$. This makes the depths of around 20 practical for emulation, and would make them affordable on hardware.

Several extensions are natural. The constraints on $T$ or on $\gamma_k$ could be added to the
warm-start optimization to avoid the long-phase layers of Sec.~\ref{sec:warm}. Warm starts, which are
derived from the SDP solution~\cite{Tate2023Quantum,Tate2023TQC}, or regularized as in
Ref.~\cite{He2026RWS}, could replace the network. Recursive schemes~\cite{Bravyi2020,
Finzgar2024} and counterdiabatic or optimized annealing schedules~\cite{Wurtz2022Counter,
Brady2021,Schiffer2024} could be combined with the annealing branch.

Finally, the noise reorders the routes. Under the Aquila error model the advantage of depth, which
drives every noiseless result above, is lost. The annealing branch is best at $p\approx5$, the
short warm start becomes the strongest protocol, and neither reaches GW in expectation or in
the probability of an optimal string per shot (Sec.~\ref{sec:noise}). 

Two features of the
noiseless picture survive. The margin of the warm start over the annealing branch, each at its best depth,
grows from $-0.008$ at $\lambda=0.1$ to $+0.076$ at $\lambda=1$. The GNN pays off most where
depth is unaffordable; and the
complementarity of the two routes persists as an allocation of shots, which removes the
blind spots of the warm start at a modest cost in shots. 

The schedules were optimized without
noise; optimizing them under noise, adding a bound on $T$ to the warm start optimization, and
mitigating the detection errors, which already restores $R_a=0.057$ for  warm start at $p=3$,
are the obvious next steps. The error budget points to the per-shot coherence and the detection
fidelity as the quantities to improve before the deep annealing branch can be tested.

\section{Summary and conclusion}
\label{sec:conclusion}

We studied an analog QAOA for weighted MaxCut problem on three-dimensional Rydberg atom arrays, in which the edge weights are written directly into three-dimensional atom positions. Unlike gadget-based embeddings or the
planar lattice graphs considered so far~\cite{Nguyen2023,Bauer2026}, this encoding requires no
auxiliary atoms. We applied it to random weighted three-regular graphs. Our emulation keeps all possible
couplings between atoms, even if they are not joined in the target graph. A fast mixer,
$\Omega_t=15.7$~rad/$\mu$s, limits the phase accumulated on these pairs. For twenty graphs with 12
vertices, we obtain the following results:

(i) The depth-one landscape has two families of maxima, and a layer-by-layer optimizer that
follows the higher one, ends on a non-adiabatic branch on which $\Ra$ saturates near $0.77$. A
single physical constraint, $\gamma_k\le1\us$, applied on every cost segment, selects an annealing
branch instead. On this branch the approximation ratio  rises monotonically to $R_a=0.959$ at $p=20$, with the same evaluation budget.

This branch exceeds the single-rounding GW expectation with $R_a=0.948$ from $p\approx16$ on; at $p=20$,
15 of 20 instances lie above it (two-sided Wilcoxon $P=3\times10^{-4}$). A global spin flip turns
the near-$\pi$ mixer rotations of this branch into small rotations of the opposite sign, and with
this identification the emulated state remains within $6\%$ of the spectral width above the ground
state of each layer's generator. The optimized analog sequence is therefore a discretized anneal,
despite the fact that its mixer always contains all pairwise atomic interactions.

(ii) A GNN warm start helps at larger depths only if the initially prepared product state is the eigenstate of
the site-resolved mixer. In the Bloqade convention, that fixes the sign of the longitudinal mixer
field. With the correct choice of this sign and $\varepsilon=0.2$, the layer-wise optimizer reaches $\Ra=0.940$ at $p=5$. For the opposite sign, $\Ra=0.895$ 
, and the transfer optimizer reaches only $0.931$ at $p=3$, with $2\us$ total laser pulse duration.

(iii) Two optimization routes fail in different instances: the warm start depends on whether the rounded
network output is optimal, and the annealing branch depends on the quality of the embedding. Keeping the
better of the two expectation values gives $\Ra=0.966$, above GW in 19 of 20 instances, while
nine variants, that merge the two routes into one pulse sequence (Appendix~\ref{app:fusion}), do
not reach this value.

(iv) Parameter transfer between the instances reduces the number of emulator calls by more than an
order of magnitude compared with the layer-wise Bayesian optimization, that makes the depth around 20 affordable. 

These statements concern the expectation values in small instances.  Both the repeated GW rounding, and the
rounded network output remain strong classical references (Secs.~\ref{sec:gw} and
\ref{sec:sampling}). 

Under the error model of the existing neutral-atom quantum processor Aquila, the advantage of
depth is lost: the best depth of the annealing branch drops from $20$ to $5$, warm start at
$p=3$ becomes the best protocol in all 20 instances (measured $\Ra=0.826$ against $0.750$), and it
yields an optimal cut in $16\%$ of the shots, less than half as often as a single GW rounding. The
noiseless ordering returns back, when all error rates are reduced tenfold, and splitting the
shots between the two routes removes the failure cases of the warm start.

Our detailed investigation of analog QAOA in three-dimensional atomic arrays shows the potential of outperforming the classical algorithms. However, that requires further improvement of the existing neutral-atom processors and reducing their noise levels.

\begin{acknowledgments}
I.B. acknowledges support from the Foundation for the Advancement of Theoretical Physics and Mathematics "BASIS".

\end{acknowledgments}

\section*{Data and code availability}
The instance files, optimized parameters, analysis scripts, the independent emulator used
for Figs.~\ref{fig:omega}--\ref{fig:embed}, and the trajectory solver and data of
Figs.~\ref{fig:noise} and~\ref{fig:noisesampling} will be made available at repository: \url{https://github.com/chenjunxi1997-cmd/QAOA-quantum-simulation-on-Bloqade.jl}.

\appendix

\section{Emulator cross-check}
\label{app:emulator}

We verified the results from Bloqade package using the emulator, which incorporates the mixer Hamiltonian, including site-resolved terms of the warm
start, as a real symmetric $4096\times4096$ matrix, and diagonalizes it once per instance and
condition. The cost segments are diagonal phases $e^{-i\gamma E_z}$ with $E_z$ being the diagonal term of
Eq.~\eqref{eq:H} at $\Omega=0$. The preparation laser pulse is propagated with a sparse matrix exponential.

For  stored optimal parameters of the annealing branch ($p=1$--$20$) and of the warm start ($p=1$--$10$, first repetition) in all 20 instances, the largest deviation from the Bloqade values is $5\times10^{-7}$. Our emulator also reproduces the $p=1$ optima of the
$\Omega_t$ scan for $\Omega_t\in\{1,2,4,8\}$ and the layer-wise baseline at $p=1$--$5$ to six
digits. 

One run is an exception (instance 20, warm start, first repetition, $p\ge4$). Here we call
recorded value the $\Ra$ that Bloqade returned for a parameter vector during the optimization
and that was saved together with it; for this run it is $0.9526$ at every $p\ge4$, the same as
at $p=3$. The emulator described here does not reproduce this value for the stored vectors at
$p\ge4$, whereas the same vectors with $\beta_4=0$ reproduce it at all $p\ge4$ to six digits.
The difference therefore comes from the fourth layer, which consists of an unusually long cost
segment ($\gamma_4=5.27\us$) followed by a mixer segment of $0.11\us$, a rotation by about
$1.7$~rad. Bloqade returns $0.9526$ for the full vector, i.e., it treats this mixer segment as if
it had no effect, while the trajectory solver of Appendix~\ref{app:noise} gives $0.574$. Bloqade
and the trajectory solver agree to $10^{-6}$ when $\gamma_4$ is set to $0$ or $1\us$, or $\beta_4$
to $0$, so the discrepancy requires the long cost segment followed by the mixer. We attribute it
to a failure of the adaptive Krylov step of the Bloqade propagator~\cite{Sidje1998} on this
segment, i.e., to the propagator used during the optimization, not to the bookkeeping of the
results.

The recorded value is attained exactly by the vector with the fourth and all later layers set to
the identity, which we use for the state-level and noise analysis. The recorded $\Ra$ values,
which enter all averages, are therefore attainable and unaffected. No other stored schedule
shows any discrepancy (Appendix~\ref{app:noise}).

\section{GW reference}
\label{app:gw}

We consider the semidefinite relaxation
\begin{align}
&\max_{\mathcal X}\ \frac14\sum_{i,j}W_{ij}\,(1-\mathcal X_{ij})\nonumber\\
&\text{subject to}\quad\mathcal X\succeq0,\ \ \mathcal X_{ii}=1,
\label{eq:sdp}
\end{align}
where $W_{ij}=w_{ij}$ for $(i,j)\in E$ and $W_{ij}=0$ otherwise, $\mathcal X$ is a real symmetric
$N\times N$ matrix and $\mathcal X\succeq0$ denotes positive semidefiniteness. It was solved with the
CLARABEL solver through CVXPY~\cite{cvxpy2016}. 

Vectors $\bm u_i$ with
$\bm u_i\cdot\bm u_j=\mathcal X_{ij}$ were obtained from the eigendecomposition of $\mathcal X$;
each rounding draws a Gaussian random vector $\bm g$ and sets $z_i=1$ if $\bm u_i\cdot\bm g>0$.
Here $2\times10^4$ roundings were drawn per instance (fixed seed). 

From the empirical distribution of the
rounded cuts we computed the expectation $\RGW$ (Monte Carlo standard error $\approx5\times10^{-4}$),
$\Pop^{\rm GW}$, and the value
\begin{equation}
\langle\max_{N_s}C\rangle=\sum_v v\,\bigl[F(v)^{N_s}-F(v^-)^{N_s}\bigr],
\label{eq:bestN}
\end{equation}
where the sum runs over the distinct cut values $v$, $F(v)$ is the probability of a cut value
$\le v$ and $F(v^-)$ that of a value $<v$. Equation~\eqref{eq:bestN} can be used as reference, because  for
independent samples, $\Pr[\max_mC(z^{(m)})\le v]=F(v)^{N_s}$. The same formula, with $F$ computed from $\Pr(z)$, gives the
QAOA best-of-$N_s$ values.

\section{Transfer optimizer}
\label{app:transfer}

\emph{Source database.} For each instance and  condition, a layer-by-layer chain is run once with
80 Gaussian-process evaluations and a 40-evaluation two-dimensional Nelder--Mead refinement per
layer. All evaluated points and the optimum of the every layer are stored. For the target instance, the
sources are the other 19 instances.

\emph{New layer.} Two new parameters are optimized by a Gaussian process with an ARD
Matérn-5/2 kernel on box-normalized inputs. The hyperparameters are refitted by maximum likelihood every
five evaluations, and the expected improvement as acquisition, is evaluated on 300 uniform and 100
local candidates. 

The initial design consists of the identity point $(0,0)$, five transferred
points, and up to 12 design points (Halton points). The transferred points are the median of the sources' optimal layers and four high-density source optima, chosen by
a Gaussian kernel density of width $0.06$ in normalized units and mutually separated by more than
$0.06$.  

The source $\gamma$ values are multiplied by
$\bar w_s/\bar w_t$, the ratio of mean edge couplings of source and target. The budget per new
layer is 30 evaluations. At $p=1$, the best point and the two best further points that are
separated from it (by more than $0.08$ in normalized units) are refined by 30 two-dimensional
Nelder--Mead evaluations each.

\emph{Refinement.} For $p\ge2$ the INTERP extrapolation~\cite{Zhou2020} of the previous optimum
is evaluated. It replaces the new-layer candidate, if it is better. Two block-coordinate sweeps
then re-optimize each earlier layer with 30 Nelder--Mead evaluations each, and a full-dimensional
Nelder--Mead refinement with 150 evaluations closes the layer. All steps accept improvements only,
and $\Ra$ is not decreasing in $p$.

\emph{Variants.} The annealing branch bounds every $\gamma_k$ by $1\us$ in all above discussion,
including the source database. The ``portfolio'' start replaces the five transferred points by a
set, which is chosen greedily to cover the sources' optima~\cite{Nemhauser1978,Wistuba2015}. A ``three
chains'' start runs the plain chain, the portfolio and a no-transfer chain simultaneously, and keeps the best. The warm-start runs were repeated three times with different seeds; we report the mean. The repetition
standard deviation at $p=10$ has a median of $0.002$; three instances (3, 14, 17) exceed $0.01$.

\section{Effective layer generator and the adiabatic test}
\label{app:adiabatic}

\emph{Decomposition.} We drop the layer index $k$. For a mixer segment of duration $\beta$ and
angle $\theta=\Omega_t\beta$, the relation
$e^{i\varphi X_i/2}Z_ie^{-i\varphi X_i/2}=Z_i\cos\varphi+Y_i\sin\varphi$, with the Pauli operators
of Eq.~\eqref{eq:pauli}, turns Eq.~\eqref{eq:HCI} into the following:
\begin{align}
H_C^I(t)=\sum_{i<j}\frac{V_{ij}}{4}\,&\bigl(Z_i\cos\varphi+Y_i\sin\varphi\bigr)\nonumber\\
\times&\bigl(Z_j\cos\varphi+Y_j\sin\varphi\bigr),
\label{eq:HCIexp}
\end{align}
with $\varphi=\Omega_t t$, up to a constant. Its average over $t\in[0,\beta]$, i.e., over $\varphi\in[0,\theta]$, is
the Eq.~\eqref{eq:Hbar} with the coefficients from Eq.~\eqref{eq:abm}. The idealized layer
\begin{align}
U^{\rm id}&=\Pi\,e^{\,i\theta^{\rm eff}\sum_iX_i/2}\,e^{-i\beta\bar H_C}\,e^{-i\gamma H_C},\nonumber\\
\theta^{\rm eff}&=\pi-\theta ,
\label{eq:Uid}
\end{align}
was compared with the exact layer for the same input state and for the state reached by the exact circuit
after the preceding layers.  The fidelity is defined as $|\langle\psi|U^{\rm id\,\dagger}U^{\rm ex}|\psi\rangle|^2$,
where $U^{\rm ex}$ is the exact layer and $|\psi\rangle$ is the input state. 

Over all layers of the annealing branch ($p=5$, $10$, $20$, 20 instances) the fidelity
lies between $0.9978$ and $1.0000$ (median $0.9989$). For  layer-wise BO schedules the
smallest value of fidelity is $0.9945$. Replacement of $\bar H_C$ by a static $H_C$, split symmetrically around the
rotation, lowers the median value to $0.90$.

\emph{Branch.} The rotation angle is reduced to the principal branch,
$\theta'=\theta-\ell\pi\in(-\pi/2,\pi/2]$ with integer $\ell$, and $\Pi^\ell$ is factored out. The
generator of Eq.~\eqref{eq:Heff} then reads $\tfrac{\theta'}{2}\sum_iX_i+\beta\bar H_C+\gamma H_C$.
In the annealing branch   for every non-trivial layer $\ell=1$. Therefore,  $\theta'=-\theta^{\rm eff}$. The layers with negligible  generators $\Omega_t(\gamma+\beta)+|\theta'|<0.05$ (all quantities dimensionless
or in rad), are excluded from the statistics. This is important  only in the instance~6 at $p=20$, where the chain appended near-identity layers 17--20 after reaching
$p=16$.

\emph{Tracking measures.} As the preparation laser pulse acts while $H_C$ is on, the
initial state is not exactly the eigenstate of $\Pi$ ($\langle\Pi\rangle=0.86$--$0.90$). The value $\langle\Pi\rangle$
is conserved afterwards, since the every later segment commutes with $\Pi$. Let $|\psi_k\rangle$ be the
emulated state after layer $k$, $w_{\rm even}=(1+\langle\psi_k|\Pi|\psi_k\rangle)/2$ the weight of its
$\Pi$-even component, and $|\phi_0^{(k)}\rangle$ the ground state of $H_k^{\rm eff}$ in the $\Pi$-even
sector. We calculate
\begin{align}
F_k&=\frac{|\langle\phi_0^{(k)}|\psi_k\rangle|^2}{w_{\rm even}},\nonumber\\
\epsilon_k&=\frac{\langle\psi_k|H_k^{\rm eff}|\psi_k\rangle-E^{(k)}_{\min}}{E^{(k)}_{\max}-E^{(k)}_{\min}},
\label{eq:Feps}
\end{align}
where $E^{(k)}_{\min}$ and $E^{(k)}_{\max}$ are the smallest and largest eigenvalues of
$H_k^{\rm eff}$ in the full space.
All eigenvalues were obtained by Lanczos iteration on the sparse $4096\times4096$ matrices.

\section{Single-protocol fusions of the two routes}
\label{app:fusion}

We attempted to join  the advantages of the annealing branch and of the warm start within one pulse sequence, instead of using the two-member portfolio of Sec.~\ref{sec:gw}. All test
runs use $\Omega_t=15.7$, the transfer optimizer of Appendix~\ref{app:transfer} with INTERP
(unless noted), one repetition, and depths $p=1$--$10$. With $c_i=2\sqrt{p_i(1-p_i)}$ and
$q_i=2p_i-1$. We considered the following variants.

\emph{(i), (ii) Warm start with and without the annealing bound.} Preparation and mixer of
Eqs.~\eqref{eq:prep} and \eqref{eq:mix} with $s=-1$ and $\varepsilon=0$; in (ii) every cost
segment is bounded by $\gamma_k\le1\us$, exactly as for the cold annealing branch.

\emph{(iii)--(v) The same bound with the $s=+1$ mixer} and $\varepsilon=0$, $0.15$, $0.3$. These
runs use the sign $s=+1$ of Sec.~\ref{sec:warm} and are included for comparison.

\emph{(vi) Decaying bias.} The mixer has the full cold-start amplitude, $\Omega_i=\Omega_t$, and a
longitudinal GNN field that decays with the layer index,
$\Delta_i\to\Delta_i-\lambda_k\Omega_t q_i$, $\lambda_k=\lambda_0\xi^{k-1}$, with initial bias strength
$\lambda_0=0.5$ and decay factor $\xi=0.7$. It means that the last layers approach the standard mixer. The initial state is the single-atom eigenstate of the first mixer, i.e., a product state with softened marginals
$p_i^{\rm eff}$ given by $2p_i^{\rm eff}-1=\lambda_0q_i/(1+\lambda_0^2q_i^2)^{1/2}$,
and $\gamma_k\le1\us$. This is a discrete analog of annealing with a decaying longitudinal bias
field built from a classical guess.

\emph{(vii), (vii$'$) Mixer ramp in $k/p$.} The initial state is the full warm state
($\varepsilon=0$), and the mixer interpolates linearly from the custom mixer to the standard one,
$\Omega_i=\Omega_t[(1-u_k)c_i+u_k]$, $\Delta_i\to\Delta_i-(1-u_k)\Omega_tq_i$, with ramp
parameter $u_k=(k-1)/(p-1)$. The first layer is the eigenstate-consistent warm start, the last layer the
standard mixer, and no parameter is added. In (vii$'$) the cost segments are also bounded by
$1\us$.

\emph{(viii) Mixer ramp in $k$.} Similar to (vii), but with $u_k=1-\xi^{k-1}$, $\xi=0.7$, which depends only on the layer index; no INTERP.

At $p=1$, variants (i), (ii), (vii), (vii$'$) and (viii) coincide (all give $0.8985$), which also
serves as a consistency check of the implementation.

\begin{figure*}[t]
\centering
\includegraphics[width=0.92\textwidth]{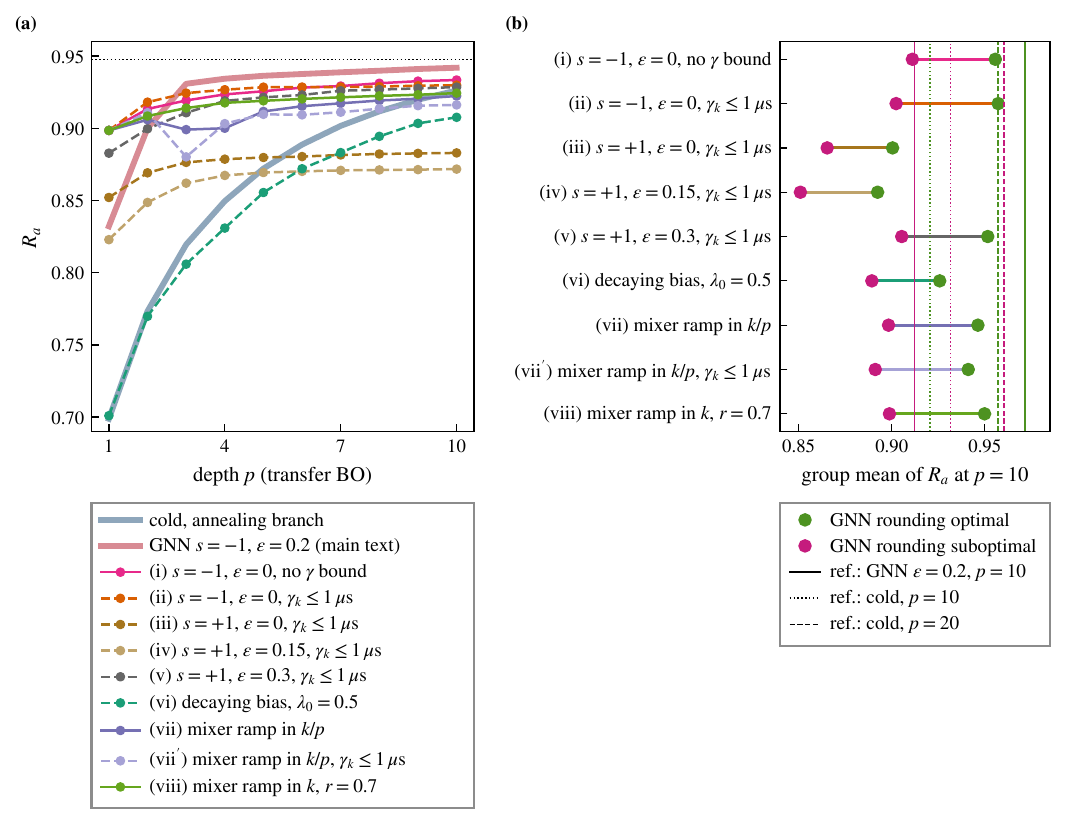}
\caption{Single-protocol merge attempts of two routes (Appendix~\ref{app:fusion}). (a) $\Ra$
as a function of depth for variants (i)--(viii) (dashed lines: $\gamma_k\le1\us$), compared with the cold
annealing branch and warm start (thick lines). The dots indicate mean $\RGW$ values.
(b) Mean $\Ra$ at $p=10$ over  10 instances with an optimal GNN rounding (green) and over the
other 10 (magenta). Vertical lines give the same group mean values for  warm start at $p=10$ (solid), and for the annealing branch at $p=10$ (dotted) and $p=20$ (dashed).}
\label{fig:fusion}
\end{figure*}

\begin{table*}[t]
\caption{Fusion variants at $p=10$ unless stated. ``Rnd.~opt.''/``Rnd.~subopt.'': mean $\Ra$ over
the instances whose GNN rounding is optimal/suboptimal. $\delta_{\rm GNN}$: mean paired
difference to the warm start of the main text ($s=-1$, $\varepsilon=0.2$, $p=10$) with
two-sided Wilcoxon $P$. Port.: mean of $\max(\Ra,\Ra^{\rm Ann,20})$. Evals: cumulative
evaluations per instance up to $p=10$. Ann.: annealing branch.}
\label{tab:fusion}
\centering
\footnotesize
\setlength{\tabcolsep}{2.6pt}
\begin{tabular}{lccccccccc}
\toprule
Variant & $p{=}1$ & $p{=}3$ & $p{=}5$ & $p{=}10$ & Rnd.\ opt. & Rnd.\ subopt. & $\delta_{\rm GNN}$ ($P$) & Port. & Evals \\
\midrule
(i) & 0.8985 & 0.9192 & 0.9256 & 0.9334 & 0.9556 & 0.9111 & $-0.0085$ ($0.044$) & 0.9605 & 4574 \\
(ii) & 0.8985 & 0.9243 & 0.9284 & 0.9298 & 0.9572 & 0.9024 & $-0.0121$ ($0.0064$) & 0.9624 & 3617 \\
(iii) & 0.8521 & 0.8763 & 0.8797 & 0.8829 & 0.9005 & 0.8653 & $-0.0590$ ($1.9\times10^{-6}$) & 0.9587 & 3924 \\
(iv) & 0.8228 & 0.8620 & 0.8693 & 0.8717 & 0.8924 & 0.8510 & $-0.0702$ ($1.9\times10^{-6}$) & 0.9587 & 4269 \\
(v) & 0.8827 & 0.9107 & 0.9214 & 0.9285 & 0.9516 & 0.9054 & $-0.0134$ ($0.002$) & 0.9601 & 4260 \\
(vi) & 0.7010 & 0.8059 & 0.8555 & 0.9076 & 0.9258 & 0.8893 & $-0.0343$ ($1.9\times10^{-5}$) & 0.9587 & 4514 \\
(vii) & 0.8985 & 0.8990 & 0.9116 & 0.9223 & 0.9463 & 0.8983 & $-0.0196$ ($8.2\times10^{-5}$) & 0.9598 & 4579 \\
(vii$'$) & 0.8985 & 0.8801 & 0.9096 & 0.9161 & 0.9411 & 0.8912 & $-0.0257$ ($5.7\times10^{-6}$) & 0.9587 & 4143 \\
(viii) & 0.8985 & 0.9140 & 0.9190 & 0.9243 & 0.9499 & 0.8988 & $-0.0176$ ($4.8\times10^{-4}$) & 0.9598 & 4568 \\
\midrule
Warm start, main text & 0.8321 & 0.9308 & 0.9363 & 0.9419 & 0.9716 & 0.9122 & --- & 0.9660 & 4554 \\
Ann.\ branch & 0.6988 & 0.8195 & 0.8721 & 0.9261 & 0.9205 & 0.9317 & --- & --- & 4599 \\
Ann.\ branch, $p{=}20$ & & & & 0.9587 & 0.9570 & 0.9604 & & & 15101 \\
\bottomrule
\end{tabular}

\end{table*}

Figure~\ref{fig:fusion}(a) and Table~\ref{tab:fusion} summarize the outcome. Every variant ends
below the warm start of the main text at $p=10$, by $0.009$ for (i) ($P=0.04$) and by up to
$0.07$ for the $s=+1$ runs. The group mean values in Fig.~\ref{fig:fusion}(b) show, why none of them
can replace the portfolio. 

In 10 instances with a suboptimal GNN rounding all variants remained
between $0.851$ and $0.911$, below the cold annealing branch at the same depth ($0.932$) and far
below it at $p=20$ ($0.960$). The bias that the GNN introduces is not removed, its
softening by clipping decays with the layer index, or is switched off gradually in the mixer. In
the 10 instances with an optimal rounding,  part of the warm-start advantage is lost
($0.893$--$0.957$ against $0.972$). 
In a portfolio with the annealing branch at $p=20$ the best of them gives $0.962$, below the $0.966$ obtained with the warm start of the main text.

Three mechanisms were observed in the runs. First, the bound $\gamma_k\le1\us$ does not turn a warm
start into an annealing path. With it, the warm-start chains flatten after a few layers [(ii):
$0.928$ at $p=5$, $0.930$ at $p=10$]. The INTERP, which the cold annealing branch accepts in 17 to
20 of the 20 instances for every $p\ge6$, is accepted in at most 2 of 20 instances at any depth in
(ii), and never in (iii) and (iv). The warm-start schedules are not smooth in $k/p$. 

Second, softening the initial state trades the head start for nothing in return: with $\lambda_0=0.5$ the
mean $|2p^{\rm eff}_i-1|$ falls from $0.68$ to $0.32$, variant (vi) starts at the cold-start value
($0.701$ at $p=1$), and the remaining bias in the mixers still keeps it below the cold branch at
$p=10$ ($-0.019$, $P=0.04$).

Third, a mixer that depends on the total depth $p$, as in (vii), breaks the property that a zero-duration layer is the identity on the depth-$(p-1)$ circuit:
adding a layer changes the Hamiltonian of every earlier mixer. Therefore, the optimized prefix is no
longer optimal. The target chains become non-monotone [(vii): $0.906$ at $p=2$, $0.899$ at
$p=3$], the source chains used for transfer degrade with depth (the annealing-bounded source
chains of (vii$'$) fall from $0.899$ at $p=1$ to $0.754$ at $p=10$ on average), and INTERP is never
accepted. Variant (viii) avoids this by letting $u_k$ depend on $k$ only; it is monotone but
converges to an intermediate value ($0.924$). 

We conclude that, at least within this family of
constructions and with the transfer optimizer, the cold annealing branch and the warm start are
best kept as separate protocols and combined only at the level of their measured expectation
values.

\begin{table}[t]
\caption{Edges of instance~1 [Fig.~\ref{fig:schematic}(a)]: atom pair $(i,j)$, distance $r_{ij}$ ($\mu$m) and edge weight $w_{ij}=V_{ij}/4$ (rad/$\mu$s). The coordinates of all atoms of the twenty instances are given in the data files accompanying this paper.}
\label{tab:inst1}
\centering
\footnotesize
\begin{tabular}{ccc@{\hspace{1.8em}}ccc}
\toprule
$(i,j)$ & $r_{ij}$ & $w_{ij}$ & $(i,j)$ & $r_{ij}$ & $w_{ij}$ \\
\midrule
(1,3) & 10.86 & 0.826 & (4,5) & 11.46 & 0.597 \\
(1,7) & 11.70 & 0.527 & (4,9) & 11.51 & 0.583 \\
(1,12) & 11.31 & 0.646 & (5,11) & 10.71 & 0.897 \\
(2,7) & 12.23 & 0.405 & (5,12) & 11.09 & 0.729 \\
(2,8) & 10.98 & 0.775 & (6,10) & 10.75 & 0.877 \\
(2,11) & 10.79 & 0.857 & (6,11) & 11.81 & 0.499 \\
(3,4) & 12.07 & 0.439 & (7,12) & 11.57 & 0.565 \\
(3,6) & 10.78 & 0.862 & (8,9) & 12.16 & 0.420 \\
(8,10) & 11.72 & 0.523 & (9,10) & 11.48 & 0.592 \\
\bottomrule
\end{tabular}
\end{table}

\begin{table*}[t]
\caption{Per-instance data at $\Omega_t=15.7$. $\eta$: unwanted-coupling ratio;
$R_{\rm rnd}$: cut ratio of the rounded GNN output; GS: target cut ratio of the ground state of
Eq.~\eqref{eq:HC} where it differs from 1; GW: single-rounding expectation; BO: layer-wise
baseline, cold start, $p=5$; Ann.: annealing branch; GNN: warm start ($s=-1$,
$\varepsilon=0.2$, transfer optimizer, mean of three repetitions); Port.: $\max$(Ann.~$p{=}20$,
GNN $p{=}10$). $\Pop$ is computed from the optimized states (warm start: first repetition).}
\label{tab:instances}
\centering
\footnotesize
\setlength{\tabcolsep}{3.6pt}
\begin{tabular}{rrrrrrrrrrrrrrr}
\toprule
 & & & & & & & BO & \multicolumn{2}{c}{Ann.} & \multicolumn{2}{c}{GNN} & & \multicolumn{2}{c}{$\Pop$} \\
\cmidrule(lr){9-10}\cmidrule(lr){11-12}\cmidrule(lr){14-15}
 & seed & $C_{\max}$ & $\eta$ & $R_{\rm rnd}$ & GS & GW & $p{=}5$ & $p{=}10$ & $p{=}20$ & $p{=}3$ & $p{=}10$ & Port. & Ann.\,20 & GNN\,10 \\
\midrule
1 & 7 & 10.094 & 0.229 & 1.000 &  & 0.9300 & 0.8409 & 0.9026 & 0.9429 & 0.9514 & 0.9626 & 0.9626 & 0.55 & 0.77 \\
2 & 11 & 11.173 & 0.072 & 1.000 &  & 0.9487 & 0.7570 & 0.9220 & 0.9729 & 0.9661 & 0.9752 & 0.9752 & 0.83 & 0.79 \\
3 & 13 & 10.419 & 0.087 & 0.813 &  & 0.9405 & 0.8372 & 0.9421 & 0.9547 & 0.8852 & 0.9082 & 0.9547 & 0.33 & 0.25 \\
4 & 15 & 11.018 & 0.084 & 1.000 &  & 0.9580 & 0.8117 & 0.9266 & 0.9775 & 0.9723 & 0.9814 & 0.9814 & 0.89 & 0.89 \\
5 & 16 & 9.760 & 0.087 & 0.921 &  & 0.9385 & 0.8626 & 0.9410 & 0.9693 & 0.9168 & 0.9453 & 0.9693 & 0.44 & 0.56 \\
6 & 17 & 9.856 & 0.276 & 0.963 & 0.973 & 0.9511 & 0.8118 & 0.9230 & 0.9541 & 0.9431 & 0.9535 & 0.9541 & 0.36 & 0.37 \\
7 & 18 & 10.372 & 0.085 & 0.940 &  & 0.9413 & 0.8514 & 0.9367 & 0.9583 & 0.9192 & 0.9244 & 0.9583 & 0.23 & 0.02 \\
8 & 19 & 10.605 & 0.056 & 1.000 &  & 0.9568 & 0.8265 & 0.9393 & 0.9721 & 0.9736 & 0.9802 & 0.9802 & 0.65 & 0.82 \\
9 & 20 & 9.966 & 0.224 & 1.000 &  & 0.9436 & 0.7405 & 0.9126 & 0.9418 & 0.9657 & 0.9741 & 0.9741 & 0.44 & 0.82 \\
10 & 22 & 10.281 & 0.112 & 0.877 &  & 0.9380 & 0.8591 & 0.9382 & 0.9544 & 0.8844 & 0.8959 & 0.9544 & 0.09 & 0.00 \\
11 & 26 & 11.250 & 0.194 & 1.000 &  & 0.9658 & 0.8514 & 0.9108 & 0.9632 & 0.9529 & 0.9678 & 0.9678 & 0.37 & 0.64 \\
12 & 31 & 10.395 & 0.223 & 0.904 & 0.978 & 0.9418 & 0.8507 & 0.9030 & 0.9470 & 0.8723 & 0.8754 & 0.9470 & 0.31 & 0.00 \\
13 & 32 & 10.654 & 0.092 & 1.000 &  & 0.9441 & 0.8437 & 0.9251 & 0.9509 & 0.9573 & 0.9636 & 0.9636 & 0.46 & 0.78 \\
14 & 34 & 11.114 & 0.071 & 0.920 &  & 0.9704 & 0.9025 & 0.9463 & 0.9660 & 0.8814 & 0.8938 & 0.9660 & 0.40 & 0.01 \\
15 & 36 & 10.629 & 0.181 & 1.000 &  & 0.9427 & 0.8483 & 0.9055 & 0.9409 & 0.9657 & 0.9749 & 0.9749 & 0.54 & 0.85 \\
16 & 41 & 10.162 & 0.084 & 0.942 & 0.990 & 0.9556 & 0.8488 & 0.9351 & 0.9700 & 0.9363 & 0.9458 & 0.9700 & 0.43 & 0.61 \\
17 & 43 & 10.664 & 0.107 & 0.893 &  & 0.9325 & 0.8520 & 0.9236 & 0.9618 & 0.8620 & 0.8706 & 0.9618 & 0.53 & 0.01 \\
18 & 47 & 11.719 & 0.111 & 1.000 &  & 0.9528 & 0.8012 & 0.9367 & 0.9600 & 0.9609 & 0.9751 & 0.9751 & 0.51 & 0.86 \\
19 & 48 & 9.448 & 0.140 & 0.937 &  & 0.9531 & 0.8703 & 0.9277 & 0.9682 & 0.8964 & 0.9088 & 0.9682 & 0.75 & 0.01 \\
20 & 51 & 10.284 & 0.177 & 1.000 &  & 0.9486 & 0.8751 & 0.9243 & 0.9477 & 0.9526 & 0.9610 & 0.9610 & 0.18 & 0.57 \\
\midrule
mean & & & 0.135 & 0.955 & & 0.9477 & 0.8371 & 0.9261 & 0.9587 & 0.9308 & 0.9419 & 0.9660 & 0.46 & 0.48 \\
\bottomrule
\end{tabular}

\end{table*}

\section{Noise model and trajectory solver}
\label{app:noise}

\paragraph{Noisy Hamiltonian.} For each shot $\zeta$ we draw independent Gaussian variables, and
make the following replacement in Eq.~\eqref{eq:H} and  all segments of the sequence:

\begin{align}
\Omega_i&\to\Omega_i\,(1+\delta_{\zeta}+\delta_{\zeta,i}),\nonumber\\
\Delta_i&\to\Delta_i+\epsilon_{\zeta}+\epsilon_{\zeta,i},\nonumber\\
\bm r_i&\to\bm r_i+\bm\rho_{\zeta,i}.
\label{eq:qs}
\end{align}
Here $\delta_\zeta$ ($\delta_{\zeta,i}$) is the global (per-atom) relative Rabi error,
$\epsilon_\zeta$ ($\epsilon_{\zeta,i}$) the global (per-atom) detuning offset and
$\bm\rho_{\zeta,i}$ the displacement of atom $i$, with the standard deviations of
Sec.~\ref{sec:noisemodel}. 

The interactions $V_{ij}$ are computed from the displaced positions, and
the compensation detuning of Eq.~\eqref{eq:comp} is taken from the nominal values. For a given draw, the
density matrix $\rho$ evolves under the master equation
\begin{equation}
\dot\rho=-i[H_\zeta,\rho]+\sum_{\ell}\Bigl(L_\ell\rho L_\ell^\dagger-\tfrac12\{L_\ell^\dagger L_\ell,\rho\}\Bigr),
\label{eq:lindblad}
\end{equation}
where $H_\zeta$ is Eq.~\eqref{eq:H} with the replacement form Eq.~\eqref{eq:qs}. Here $\{\cdot,\cdot\}$ is the
anticommutator, and the jump operators are
$\{L_\ell\}=\{\sqrt{\gamma_\phi}\,Z_i,\ \sqrt{\gamma_1}\,|g\rangle\langle r|_i\}_{i=1}^{N}$.

\paragraph{Trajectories.} Since $Z_i^\dagger Z_i$ is the identity, the dephasing jumps occur at a
constant total rate $N\gamma_\phi$ independent of the state. The time and the number of the atom, for which the dephasing is applied, are drawn in
advance as a Poisson process. The decay is treated with the waiting-time
algorithm~\cite{Dalibard1992,Molmer1993}: the  state evolves under the non-Hermitian Hamiltonian
$H_{\rm eff}=H_\zeta-\tfrac{i}{2}\gamma_1\sum_in_i$ until its squared norm falls below a uniform
random number. 

The jump time is located by bisection, and the decaying atom is chosen with probability proportional to $\langle n_i\rangle$. The cost segments are diagonal, and are propagated exactly. The preparation and mixer segments are propagated with an Arnoldi method (Krylov dimension
20, Gram--Schmidt with re-orthogonalization) with time step halved, until the a-posteriori
error estimate falls below $10^{-10}$, relative to the norm of the state. 

The measured bit-string
distribution is the trajectory average of $|\langle z|\psi\rangle|^2$ after the detection
channel. It is a tensor product of the single-atom matrix
$\bigl(\begin{smallmatrix}1-\pi_{01}&\pi_{10}\\ \pi_{01}&1-\pi_{10}\end{smallmatrix}\bigr)$, where
$\pi_{01}$ ($\pi_{10}$) is the probability of reading $|g\rangle$ as $|r\rangle$ ($|r\rangle$ as
$|g\rangle$). From it we compute $\Ra$, $\Pop$ and the best-of-$N_s$ statistics as in
Sec.~\ref{sec:fom}. The probability of finding an optimal cut among $N_s$ independent shots is
$1-(1-\Pop)^{N_s}$, and $1-(1-\Pop^{(a)})^{N_s/2}(1-\Pop^{(b)})^{N_s/2}$ when the shots are
split equally between two protocols $a$ and $b$.

\paragraph{Validation.} (i) Without noise the solver reproduces Bloqade for random parameter
vectors (three instances, $p=1$ and $3$, both routes) to $1.4\times10^{-11}$, and the stored
optimal values of the evaluated schedules to $2\times10^{-10}$, except for  two vectors of
instance 20, already discussed in Appendix~\ref{app:emulator}. (ii) For a non-Hermitian $H_{\rm eff}$,
on eight atoms the Krylov propagator agrees with the dense matrix exponential to
$8\times10^{-13}$. (iii) On four atoms, with rates enlarged to $\gamma_1=0.4$ and
$\gamma_\phi=0.3\,\mu\mathrm{s}^{-1}$, the average of $3000$ trajectories agrees with the exact
solution of Eq.~\eqref{eq:lindblad} within its statistical error ($z$ scores $+0.45$ for the
cold start and $-0.80$ for the warm start). (iv) The detection channel and the best-of-$N_s$
formula were checked against explicit enumeration and Monte Carlo sampling. With 300
trajectories the statistical error of an instance average is below $0.002$ in $\Ra$.

\section{Per-instance results}
\label{app:table}

Table~\ref{tab:inst1} lists the edges of instance~1, and Table~\ref{tab:instances} presents the
results for all twenty instances.

\bibliographystyle{quantum}

\end{document}